\documentclass[namedreferences,hyperref,optionalrh,solaromanenum]{spr-sola}

\usepackage{graphicx}                    % For eps figures, newer & more powerfull
\usepackage{color}                       % For color text: \color command
\newcommand{\apj}{{\it Astrophys. J.}}
\newcommand{\apjl}{{\it Astrophys. J. Lett.}}

\newcommand{\pasj}{{\it Publ. Astron. Soc. Japan}}

\chardef\us=`\_

\newcommand{\mml}{mm/submm-$\lambda$}

\begin{document}
%\begin{article}

%\begin{opening}
\begin{frontmatter}
\title{Observing the Sun with the Atacama Large Millimeter/submillimeter Array (ALMA): Polarization Observations at 3 mm}

%%%%%%%%%%%%%%%%%%%%%%%%%%%%%%%%%%%%%%%%%%%%%%%%%%%
%% Authors Names
%
% \author[addressref={},corref,email={}]{\inits{}\fnm{}\lnm{}\orcid{}}
\author[addressref={aff1,aff2},corref,email={masumi.shimojo@nao.ac.jp}]{\inits{M.}\fnm{Masumi}~\lnm{Shimojo}\orcid{0000-0002-2350-3749}}
\author[addressref={aff3},corref,email={tbastian@nrao.edu}]{\inits{T.S.}\fnm{Timothy S.}~\lnm{Bastian}\orcid{0000-0002-0713-0604}}
\author[addressref={aff4,aff1,aff2},corref,email={seiji.kameno@nao.ac.jp}]{\inits{S.}\fnm{Seiji}~\lnm{Kameno}\orcid{0000-0002-5158-0063}}
\author[addressref={aff3},corref,email={ahales@nrao.edu}]{\inits{A.S.}\fnm{Antonio S.}~\lnm{Hales}\orcid{0000-0001-5073-2849}}

%%%%%%%%%%%%%%%%%%%%%%%%%%%%%%%%%%%%%%%%%%%%%%%%%%%
%% Runningheads
%
\runningauthor{Shimojo et al.}
\runningtitle{Solar Polarimetry at 3mm}

%%%%%%%%%%%%%%%%%%%%%%%%%%%%%%%%%%%%%%%%%%%%%%%%%%%
%% Affilations 
%% id shold be the same with \author addressref value.
%\address[id={}]{}
\address[id={aff1}]{National Astronomical Observatory of Japan, Mitaka, Japan}
\address[id={aff2}]{Graduate University for Advanced Studies, SOKENDAI, Mitaka, Japan}
\address[id={aff3}]{National Radio Astronomy Observatory, Charlottesville, VA, United States}\
\address[id={aff4}]{Joint ALMA Observatory, Santiago, Chile}

%%%%%%%%%%%%%%%%%%%%%%%%%%%%%%%%%%%%%%%%%%%%%%%%%%%
%%% Abstract 
%\begin{abstract}
%\end{abstract}
\begin{abstract}
The Atacama Large Millimeter-submillimeter Array (ALMA) is a general purpose telescope that performs a broad program of astrophysical observations. Beginning in late-2016, solar observations with ALMA became available, thereby opening a new window onto solar physics. Since then, the number of solar observing capabilities has increased substantially but polarimetric observations, a community priority, have not been available. Weakly circularly polarized emission is expected from the chromosphere where magnetic fields are strong. Hence, maps of Stokes~V provide critical new constraints on the longitudinal component of the chromospheric magnetic field. Between 2019-2022, an ALMA solar development effort dedicated to making solar polarimetry at millimeter wavelengths a reality was carried out. Here, we discuss the development effort to enable solar polarimetry in the 3~mm band (ALMA Band 3) in detail and present a number of results that emerge from the development program. These include tests that validate polarization calibration, including evaluation of instrumental polarization: both antenna based ``leakage" terms and off-axis effects (termed ``beam squint" for Stokes V). We also present test polarimetric observations of a magnetized source on the Sun, the following sunspot in a solar active region, which shows a significant Stokes V signature in line with expectations. Finally, we provide some cautions and guidance to users contemplating the use of polarization observations with ALMA. 
\end{abstract}

%%%%%%%%%%%%%%%%%%%%%%%%%%%%%%%%%%%%%%%%%%%%%%%%%%%
%% Keywords
%

\end{frontmatter}

%-------------------------------------------------

%%%%%%%%%%%%%%%%%%%%%%%%%%%%%%%%%%%%%%%%%%%%%%%%%%%
%% Sections
%
% \section{}%\label{s:?} 

\section{Introduction}\label{sec:Intro} 

Solar observations at millimeter and submillimeter waves (\mml) offer unique diagnostics of solar phenomena, particularly the solar chromosphere. For non-flaring emission, the source function is Planckian, the Rayleigh-Jeans approximation is valid, and the observed continuum intensity is simply related to the kinetic temperature of the emitting material. Historically, however, observations at \mml\ have been confined to single dish measurements or sparse interferometric arrays with limited imaging capabilities \citep[see, e.g.,][]{2016SSRv..200....1W}. As a result, \mml\ observations were generally of low angular resolution although important progress was made on characterizing both quiet Sun and flare emissions. With the advent of the Atacama Large Millimeter/ submillimeter Array \citep[ALMA:][]{2009IEEEP..97.1463W} the state of affairs changed dramatically. ALMA provides both high angular resolution and high time resolution observations, fundamentally opening a new window onto the Sun.  The extraordinary potential of \mml\ observations was summarized by \cite{2016SSRv..200....1W}. While ALMA was dedicated in 2012, it was not until late-2016 (ALMA Cycle 4), that solar observing became available in two wavelength bands, Band 3 (3~mm) and Band 6 (1.25~mm).  Since then, additional bands and observing modes have been made available to solar observers and have been exploited by numerous observers, as presented by the collection of papers in the special issue of Frontiers in Astronomy and Space Sciences ``The Sun Seen with the Atacama Large mm and sub-mm Array (ALMA) - First Results" (2022). To date, however, only continuum observations in total intensity (Stokes~I) have been possible. A longstanding goal of the community has been to enable solar  full Stokes polarimetry. Stokes~V measurements are of particular interest as an important diagnostic of chromospheric magnetic fields. Here, we report efforts that led to full Stokes polarimetry being offered to the solar community in the 3~mm band in ALMA observing Cycle~10.

In general, incoherent radio emission from the Sun can be circularly polarized as the result of the emission mechanism or as a result of the refractive index. Thermal gyroresonance emission at low harmonics of the electron gyrofrequency is relevant at cm-$\lambda$ and the emission at these wavelengths can be highly circularly polarized, producing strong Stokes~V emission \citep{1997SoPh..174...31W}. From 17 GHz Stokes-I and V images obtained with Nobeyama RadioHeliograph \citep[NoRH:][]{1994IEEEP..82..705N}, the sources by this emission mechanism with 17 GHz above the sunspots are reported by \cite{1994PASJ...46L..17S} and \cite{2006PASJ...58...11V}.  In addition, a bright and point-like source above a sunspot without any flares in the 34 GHz Stoke-I images is sometimes observed by the NoRH. \cite{2019ApJ...880L..29A} and \cite{2023ApJ...943..160F} have claimed that such sources are due to thermal gyroresonance emission. However, it is hard to confirm the emission mechanism because the 34 GHz channel of NoRH does not measure Stokes-V. Gyroresonance emission is not believed to be relevant at frequencies over 100 GHz because magnetic fields of order 12 kG in the chromosphere and corona would be required to render thermal plasma optically thick by this mechanism. On the other hand, non-thermal electrons accelerated in a solar flare emit gyrosynchrotron radiation at higher harmonics of the gyro-frequency, extending to \mml\ \citep{2019SoPh..294..108H}. Nonthermal gyrosynchrotron emission can be moderately cicularly polarized \citep{1969ApJ...158..753R}. Observations of solar flare emission with ALMA have not yet been commissioned, however. Nevertheless, thermal free-free emission can be weakly circularly polarized in the presence of strong magnetic fields. This is due to the fact that the refractive index of a magnetoactive plasma  results in an opacity difference between the ordinary (o) and extraordinary (x)  modes. We do not expect there to be a significantly linearly polarized component of the emission to be present (Stokes Q and U) because of the extremely high Faraday depth of the medium. Using a 3D non-LTE radiative MHD simulation of the  solar chromosphere, \citet{2017A&A...601A..43L} estimate the degree of circular polarization of thermal free-free emission at 3~mm, predicting a degree of circular polarization $V/I \sim$1.5 \% circular in an active region and showing that polarimetry with ALMA is a powerful tool for producing chromospheric magnetograms of the longitudinal field \citep{2020FrASS...7...45L}. In this paper, therefore, we address polarimetric observations of active regions in the 3~mm band (ALMA Band 3). 

A testing and commissioning program designed to enable solar polarimetry was first developed in 2019 but, owing to the pandemic, was not fully executed until 2022.  These efforts resulted in datasets obtained during the commissioning activities that have been released to the community as the Scientific Verification (SV) data from the ALMA observatory. However, the use of the solar polarization mode has some limitations that are greater than those of other solar observing modes of ALMA. In addition, the interpretation of the full-Stokes solar data is more complex. In order to reduce these difficulties and to encourage solar polarization observations with ALMA, this paper describes the characteristics of solar polarization observations and the SV data as a guide to solar polarization data obtained with ALMA.

In the following sections, we briefly summarize the solar observing with ALMA and the challenges posed by polarization observations. We then describe the test program and results for polarization observations in the 3~mm band. We explain the details of the polarization observing mode offered in ALMA Cycle~10 and provide guidance for users. Finally, we summarize this development and the prospects for solar polarization observations with ALMA.

\section{Observing the Sun with ALMA}\label{sec:SolALMA}

\subsection{Overview}\label{subsec:USO}

The Sun is a challenging target for most radio interferometers not dedicated to solar observations, including ALMA. The major reasons are:
\begin{enumerate}
\item[\#1] The Sun can be much larger than the field of view of the antennas, which is only $\approx 1'$ for ALMA 12~m antennas at a wavelength of 3~mm.
\item[\#2] The Sun is a much more intense source compared with the other celestial objects, causing saturation of the antenna electronics and a loss of coherence.
\item[\#3] Solar emission is spatially complex and dynamic on time scales as short as seconds, requiring excellent instantaneous sampling of spatial frequencies in the Fourier domain.
\end{enumerate}
To address these issues special measures must be implemented. The details for solar observations with ALMA are discussed by \cite{2017SoPh..292...87S} and \cite{2017SoPh..292...88W}. We briefly reiterate the essential features here. 

Items \#1 and \#3 are related and lead to the question: ``How can we get as many Fourier samples as possible instantaneously in the uv plane from zero spatial frequency to the limit provided by the array configuration?" The answer for ALMA solar observations involves several factors. First, solar observations are limited to the most compact array configurations to ensure high density sampling. For the 3~mm band (Band 3) observations are offered in array configurations C-1, C-2, C-3, and C-4, the largest having a maximum antenna baseline of 780~m. Second, in contrast to non-solar programs, the Sun is observed with a heterogeneous array comprising 7m and 12m antennas to increase the number of measurements of short spatial frequencies. Third, full disk total power (TP) maps of the Sun can be Sun created that are essentially simultaneoud with the interferometric observations. Full disk TP maps may be used fill in the smallest spatial frequencies not measured by the interferometric array. Depending on the science goals of a particular observing program, mosaicking techniques may also be employed to enlarge the domain mapped by the array. 

To address Item \#2 we developed the ``Mixer-Detuning" technique. \cite{2013imtw.confE...1Y} showed that we can reduce the gain of the ALMA Superconductor-Insulator-Superconductor (SIS) mixers by changing the bias voltage\footnote{In the case of ALMA Band 5 (1.51~mm) the local oscillator current is also adjusted.} from its nominal value. In particular, since the dynamic range of the receiver scales roughly inversely with gain, the voltage bias is set to ensure that stronger signal levels may be observed without saturating the receiver. We refer to this receiver mode as the ``MD-mode." While mixer-detuning reduces the receiver gain, the signal must still be attenuated so as to avoid saturation of other electronic elements along the signal path. When we optimize the attenuator levels of the receiver system using the MD-mode for the signal level of the Sun, we cannot detect the quasars or blazars typically observed as calibrators because the solar attenuation level is too high. To avoid the problem, we reduce the attenuator levels by a fixed value for all antennas when we observe calibrators. Changing the attenuator levels introduces a phase shift into the signal path. However, when the differences between the attenuator level for the Sun and the calibrators are the same for all antennas, the phase shifts effectively difference out. Hence, calibration of ALMA data is robust.

\subsection{Solar Polarimetry with ALMA}\label{subsec:APCS}

Polarization calibration for non-solar observations is described in the ALMA Technical Handbook \citep{THBCy9}, which is published annually for each ALMA observing cycle. In this section, we briefly describe polarization observations and calibration. We then discuss the challenges of solar polarization observations. 

Full Stokes polarimetry of sidereal radio sources requires measurements of two orthogonal polarizations. ALMA uses native linear antenna feeds and the two polarizations are therefore linear. To perform these measurements each frequency band cartridge in the receiver cabin of each ALMA antenna contains two complete receiver systems sensitive to orthogonal linear polarizations, X and Y. Until ALMA Cycle 10 in 2023, solar polarimetry has not been supported. Only continuum measurements of the total flux density in Stokes I were possible. Specifically, the correlator only produced correlation products XX and YY, from which Stokes-I maps can be formed. Calibration of the data in this mode is detailed by \citet{2017SoPh..292...87S}. Briefly, a given solar observation is preceded by observations of a bandpass and flux calibrator, followed by source scans interleaved with observations of a phase calibrator. From these calibrator observations the complex gain of each polarization channel X and Y can be deduced for each antenna relative to a reference antenna and applied to the source scan data. 

In order to perform Stokes polarimetry ALMA's correlator produces four cross-correlations (XX, YY, XY, and YX) from the X and Y signals for each antenna baseline. For an ideal instrument, the correlations, as measured by complex visibilities $V$ on a given antenna baseline, are related to the four Stokes parameters I, Q, U, and V as follows:

\begin{eqnarray}
V_{XX}&=&I+Q \qquad V_{YY}=I-Q \\
V_{XY}&=&U+iV \qquad V_{YX}=U-iV
\end{eqnarray}

\noindent and so the Stokes parameters can be recovered as 

\begin{eqnarray}
I&=&\frac{V_{XX}+V_{YY}}{2} \qquad Q=\frac{V_{XX}-V_{YY}}{2} \\
U&=&\frac{V_{XY}+V_{YX}}{2} \qquad V=\frac{V_{XY}-V_{YX}}{2i}
\end{eqnarray}

\noindent from which it can be seen that a measurement of circularly polarized emission requires the cross-hand correlations XY and YX. In practical terms, it is necessary to know the complex gain of each antenna for each polarization channel. For polarization measurements it is also necessary to account for the fact that the analog signal path for the X and Y polarization signals is slightly different. The difference causes a delay between the X and Y signals in the reference antenna known as the ``X-Y offset".  Moreover, the division of the incident signal into orthogonal X and Y polarization just after the feed horn at each antenna is not perfect, and crosstalk between them occurs. This is called polarization ``leakage", or instrumental polarization, and is embodied in the so-called ``D-terms", antenna-based complex coefficients that capture the leakage of one polarization channel into the orthogonal channel. Careful design has limited the magnitude of D-terms to be small: typically 1-2\% for celestial observing.  Nevertheless, polarization leakage considerably complicates calibration because the Stokes parameters are now coupled to a given correlation through leakage terms. In addition, for linearly polarized feeds, Stokes Q and U vary with parallactic angle $\psi$ as the antenna response pattern rotates on the sky. Hence, for an antenna baseline $ij$ formed by antenna $i$ and antenna $j$ we have the following correlations:

\begin{eqnarray}
V_{XX}&=&(I+Q_\psi)+(U_\psi+iV)d^\ast_{X_j}+d_{X_i}(U_\psi-iV)+d_{X_i}(I-Q_\psi)d^\ast_{X_j} \\
V_{YY}&=&d_{Y_i}(I+Q_\psi)d^\ast_{Y_j}+d_{Y_i}(U_\psi+iV)+(U_\psi-iV)d^\ast_{Y_j}+(I-Q_\psi) \\
V_{XY}&=&(I+Q_\psi)d^\ast_{Y_j}+(U_\psi+iV)+d_{X_i}(U_\psi-iV)d^\ast_{Y_j}+d_{X_i}(I-Q_\psi) \\
V_{YX}&=&d_{Y_i}(I+Q_\psi)d^\ast_{Y_j}+d_{Y_i}(U_\psi+iV)+(U_\psi-iV)D^\ast_{Y_j}+(I-Q_\psi)
\end{eqnarray}

\noindent To calibrate the D-terms and the X-Y offset bright, compact sources such as quasars or blazars with well-known polarization properties are observed as polarization calibrators. Such sources have a significantly linearly polarized component ($\ge$a few \%) with a known electric vector position angle, but they generally have no significant circularly polarized emission. To disentangle the dependencies on the parallactic angle, observations of the polarization calibrator must be carried out over a sufficient range of $\psi$. Therefore, the total observation time currently required for ALMA polarization calibration exceeds 3 hours, even if the science goals of a given program do not require a duration this long.  The polarization parameters of a given polarization calibrator also depend on the observing frequency. Hence, D-terms are measured as a function of frequency for each polarization channel.  

An issue that effects all observers interested in polarimetric observations is the nature and impact of off-axis instrumental polarization due to the asymmetric optics of the antennas. ALMA supports observing in 10 frequency bands spanning 35-950 GHz. In each antenna receiver cabin the receivers are distributed around the optical axis of the antenna; i.e., the axis of the receiver optics is offset relative to the axis of the antenna reflector. Consequently, off-axis polarization effects must be assessed and corrected as needed. In effect, the D-terms are direction-dependent. At present, polarization calibration only solves for the D-terms on-axis and hence the off-axis effects remain. Polarimetry in Stokes I, Q, and U was first offered in ALMA Cycle 7 and the impact of off-axis effects on linearly polarized sources has therefore received the most attention to date \citep[e.g.,][]{2020PASP..132i4501H}. Solar sources are not expected to be linearly polarized, in general, but detectable circularly polarized sources are expected. Hence, in this paper, we consider off-axis polarization effects on circularly polarized emission in \S3.3 as part of the testing and commissioning effort. 

Finally, although ALMA has offered polarization observations for non-solar targets since Cycle 7, such observations were initially only supported for the 12-m array. Beginning in ALMA Cycle 9 the capability was also offered for the standalone 7-m array. Our commissioning activities for solar polarization observations focused on the 12-m array and solar polarimetry is only offered on the 12-m array at present.

\subsection{Band Selection for Solar Polarimetry with ALMA}\label{subsec:BSSPO}

The procedures required to perform ALMA polarimetry in general, and solar polarimetry in particular, apply to all ALMA frequency bands. However, enabling the capability in a given frequency band requires considerable resources and a number of factors were considered in prioritizing Band~3 as the first frequency band  for development of the capability. 

On scientific grounds, lower frequency bands are favored for the detection of significant Stokes~V signals from the Sun. The degree of circular polarization resulting from thermal free-free emission in the presence of strong magnetic field depends on the reciprocal of the observing frequency. Therefore, the possibility of detecting the circular polarization signals of strong magnetic structures, such as sunspots, becomes higher when we observe at lower frequencies. We note that, beginning with ALMA Cycle 10, Band~1 (35-50~GHz) was offered for the first time. While it is certainly of interest for solar polarimetric observations, it has not yet been commissioned for solar observations of any kind and it was therefore premature to prioritize it for polarization measurements. Band~3, nominally 100~GHz ($\lambda=3$ mm) was one of the first ALMA frequency bands to be commissioned for solar observing and the community therefore has considerable experience interpreting the data. Finally, with the exception of Band~1, Band~3 has the largest field of view. Hence, Band~3 was the first to be tested and commissioned for solar polarimetry. 

\section{Testing and Commissioning Band 3 Solar Polarimetry}\label{sec:obs}

As mentioned in \S\ref{subsec:USO},  there are two key differences in performing solar observations compared to non-solar observations:

\begin{itemize}
\item Both the calibrators and the Sun are observed with the MD mode.
\item When observing calibrator sources, the attenuation levels of all antennas are reduced by a fixed equal amount from the levels for the Sun.
\end{itemize}

\noindent For polarization observations of the Sun it was necessary to verify that these changes do not significantly affect the polarization calibration or, if it does, to identify the steps to be taken to correct or mitigate such effects. Therefore, the testing and commissioning of Band 3 solar polarimetry proceeded as follows:

\begin{enumerate}
\item Evaluate whether or not the MD-mode affects polarization calibration and characterize how it affects it.
\item Evaluate the detectability of a polarized signal (Stokes V) from a magnetized solar source (a sunspot) with ALMA polarization observations.
\item Evaluate the off-axis polarization performance of the MD mode.
\end{enumerate}

\noindent Step i) is described in the next section. Step ii) and iii) are explained in \S\ref{subsec:TOSun}.

\subsection{Validation of the MD Mode}\label{subsec:MDNormO}

To evaluate the effect of the MD mode on the polarization calibration, we observed 3C279, a blazar with known polarization properties, with both the nominal and MD modes. Initial observations were made in late 2019, but due to the disruptions caused by the pandemic, further testing was suspended for a considerable period of time. It was not until 2022 that we were able to complete test observations using nominal and MD mode receiver settings under similar observing conditions. 

The successful observation was performed in the afternoon of July 31, 2022, using a 12-m array of 37 antennas.  First, we observed 3C279 for 1.5 hours using the nominal and MD-mode. We call this the ``first half". Three 12 m antennas -- DA61, DA62, and DV02\footnote{The 12m-array is assembled from ``DA'' antennas built by the AEM Consortium, Europe and ``DV'' antennas built by Vertex RSI, USA.}  -- were operated in the MD mode, and the other 34 antennas were operated in the nominal mode during the first half. Then, we observed the same target for another 1.5 hours using only nominal settings. This is called the ``second half". The total duration of the observations covers about 100 degrees of parallactic angle. The observations are performed with the standard observing sequence for polarization observations of non-solar celestial objects, except for the use of the MD mode for the three antennas during the first half.

\subsubsection{The effect of the MD mode on the X-Y offset}

\begin{figure} 
\centerline{\includegraphics[width=1\textwidth,clip=]{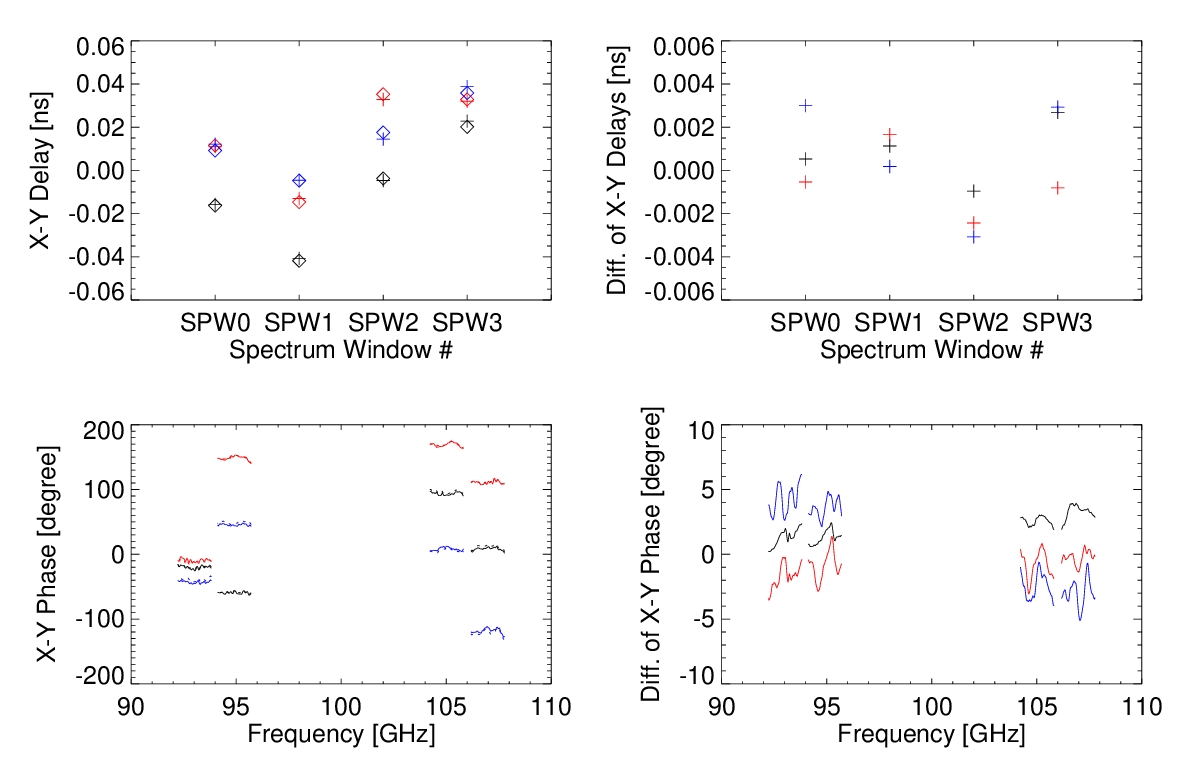}}
\caption{Upper Left: Delay between X and Y for each spectral window. The pluses and asterisks indicate the delays derived from the first and second half respectively. Upper Right: The difference of X-Y delay between them. Lower Left: Phase difference between X and Y. The solid lines show the phase differences estimated from the first half. and the dotted lines show those estimated from the second half. Lower Right: The lines show the differences between them. In these plots, Black, Red, and Blue indicate the DA64, DA62, and DV02 antennas, respectively.}\label{fig:XY_pd}
\end{figure}

Since the X-Y offset is caused by the difference between the X and Y signal paths in the reference antenna. We evaluated the X-Y offset using reference antennas operating in both the normal model and in MD mode. In particular, we used DA62 (MD mode), DA64 (normal) and DV02 (MD mode) as reference antennas. The calibration table obtained for DA64 is used as the comparison standard because the antenna was operated in normal mode during the entire observation. All three antennas were located near the center of the array, and the distances between the antennas were less than 100 m. Thus, the effect of the atmosphere would be minimized.

The X-Y offset has two components. One is a delay, and the other is a phase difference. The upper-left panel of Figure \ref{fig:XY_pd} shows the delay component of the X-Y offset for each spectral window, and the upper-right panel indicates the difference between the delays estimated from the first and second half.The average difference of the delays in each antenna is 0.0013 ns for DA62, 0.0023 ns for DV02, and 0.0013 ns for DA64. We conclude that since the difference in the delays for DA62 and DV01 are quite similar to that in DA64  the effect of the MD-mode on the delays is negligible. We also find no significant impact of the MD mode on the phase differences estimated from the first and second half, respectively, as shown in the lower-left panel of Figure \ref{fig:XY_pd}. The average difference for each antenna is 0.9 degrees for DA62, 3.3 degrees for DV02, and 2.1 degrees for DA64. We consider these values to be acceptable and conclude that use of the MD mode does not have a significant impact on the X-Y offset.

\subsubsection{The effect of the MD mode on the D-term}

\begin{figure} 
\centerline{\includegraphics[width=1\textwidth,clip=]{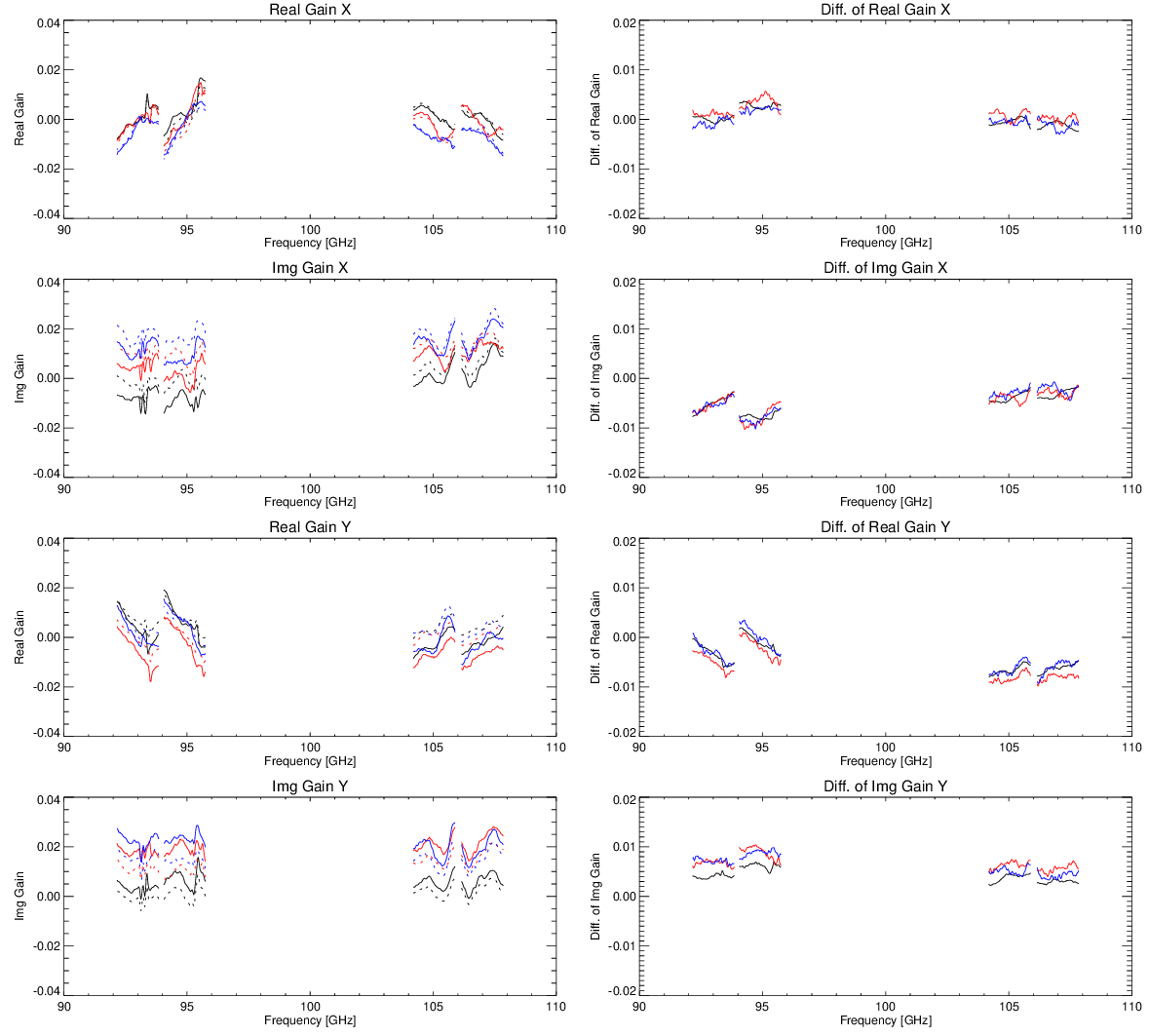}}
\caption{Left: the gain tables for calibrating the D-Term From top, Real Gain for X,  Imaginary Gain for X,  Real Gain for Y,  and  Imaginary Gain for Y.  The solid lines show the gains derived from the first half, and the dotted lines show those derived from the second half. Right: The lines show the differences between them. The color code in the plots is the same as Figure \ref{fig:XY_pd}.}\label{fig:Dterm}
\end{figure}

We have verified impacts of MD on D-terms by following scheme, and concluded that no significant change under MD, compared with MD, was observed.

The second half observation with all antennas under non-MD was used as a control. We determined D-term spectra by solving the equation
\begin{eqnarray}
\mathbf{X} = D P \mathbf{S}, \label{eqn:pol_fundamental_equation}
\end{eqnarray}

where $\mathbf{X} = (\left< X_m X^*_n\right>, \left< X_m Y^*_n\right>, \left< Y_m X^*_n\right>, \left< Y_m Y^*_n\right> )^{T}$ is the gain-calibrated visibilities of antenna pair $m$ and $n$, and $\mathbf{S} = (I, Q, U, V)^{T}$ is the Stokes parameters of the calibrator.
$P$ is the matrix of parallactic angle $\psi$ described as
\begin{eqnarray}
P = 
\left(
\begin{array}{cccc}
1 &  \cos 2\psi  &  \sin 2\psi & 0 \\
0 & -\sin 2\psi  &  \cos 2\psi & i \\
0 & -\sin 2\psi  &  \cos 2\psi & -i \\
1 & -\cos 2\psi  & -\sin 2\psi & 0 \\
\end{array}
\right), \nonumber
\end{eqnarray}
and $D$ is the D-term matrix,
\begin{eqnarray}
D = 
\left(
\begin{array}{cccc}
1              & D^{n*}_X        & D^m_X           & D^m_X D^{n*}_X \\
D^{n*}_Y       & 1               & D^m_X D^{n*}_Y  & D^m_X          \\
D^m_Y          & D^m_Y D^{n*}_X  & 1               & D^{n*}_X       \\
D^m_Y D^{n*}_Y & D^m_Y           & D^{n*}_Y        & 1              \\
\end{array}
\right). \nonumber
\end{eqnarray}

Giving calibrated visibilities and the initial Stokes parameters estimated by AMAPOLA, we have tentative solutions for D-term spectra.
They are used to improve Stokes parameters by applying equation \ref{eqn:pol_fundamental_equation}.
After two iteration cycles, we obtain solutions for the Stokes parameters and the D-term spectra.

For the first half execution where we set MD on DA62, DV02, and DA64, we derived their D-terms by using D-term transfer method.
D-terms of non-MD antennas were estimated in the same manner described above.
Now, equation \ref{eqn:pol_fundamental_equation} containing unknown D-terms of MD antenna, $m$, are 
\begin{eqnarray}
\left< X_m X^*_n\right> + \left< X_m Y^*_n\right> &=& ((1 + D^{n*}_Y) PS_0 + (1 + D^{n*}_X)PS_1) + \left[ (1 + D^{n*}_Y) PS_2 + (1 + D^{n*}_X)PS_3 \right] D^m_X, \nonumber \\ 
\left< Y_m Y^*_n\right> + \left< Y_m X^*_n\right> &=& ((1 + D^{n*}_Y) PS_2 + (1 + D^{n*}_X)PS_3) + \left[ (1 + D^{n*}_Y) PS_0 + (1 + D^{n*}_X)PS_1 \right] D^m_Y, \label{eqn:unknownD}
\end{eqnarray}
where $(PS_0, PS_1, PS_2, PS_3) = (I + Q \cos 2\psi + U \sin 2\psi, U \cos 2\psi - Q \sin 2\psi + iV, U \cos 2\psi - Q \sin 2\psi - iV, I - Q \cos 2\psi - U \sin 2\psi)$.

Solving equation \ref{eqn:unknownD} for $D^m_X$ and $D^m_Y$, we have
\begin{eqnarray}
D^m_X &=& \frac{\sum_n \left< X_m X^*_n\right> + \left< X_m Y^*_n\right> - ((1 + D^{n*}_Y) PS_0 + (1 + D^{n*}_X)PS_1) }{\sum_n (1 + D^{n*}_Y) PS_2 + (1 + D^{n*}_X)PS_3}, \nonumber \\
D^m_Y &=& \frac{\sum_n \left< Y_m Y^*_n\right> + \left< Y_m X^*_n\right> - ((1 + D^{n*}_Y) PS_2 + (1 + D^{n*}_X)PS_3) }{\sum_n (1 + D^{n*}_Y) PS_0 + (1 + D^{n*}_X)PS_1}.
\end{eqnarray}

Figure 2 shows the calibration gains for the D-term (Left panels) and the differences between the first and second halves (Right panels). As shown in the right panels of the Figure, the difference between the first and second halves of DA62 and DV02 is similar to that of DA64, and the differences are smaller than the gain of the D-term itself. Hence, we found from the test observations that the effect of MD mode on the D-terms is negligible.
We conclude that the MD mode has no significant impact on determination of the X-Y offset or on D-terms for observations of calibrator sources. We then proceeded to perform polarimetric test observations of a solar target.

\subsection{Test Observations of Solar Sources}\label{subsec:TOSun}

\subsubsection{Observation and Calibration}

After confirming that the effect of the MD-mode on polarimetric observations is negligible for a known calibrator source, we needed to demonstrate the detection of Stokes-V signals from the Sun and to evaluate the off-axis instrumental polarization. However, as described in \S\ref{subsec:USO}, solar observations are limited to compact array configurations. Therefore, even before we had completed our assessment of the polarimetry of 3C279 using the MD-mode, we carried out a solar polarization observation for a sunspot with a single-pointing on 14 May 2022. Unfortunately, due to the telescope time limitations , the target was observed for only 1.5 hours. The total observation time was insufficient to fully calibrate the D-terms, and the precision of the polarization measurement was therefore degraded. Nevertheless, a preliminary detection of Stokes-V signals above the sunspot were made at a 7$\sigma$ level. 

\begin{figure} 
\centerline{\includegraphics[width=1\textwidth,clip=]{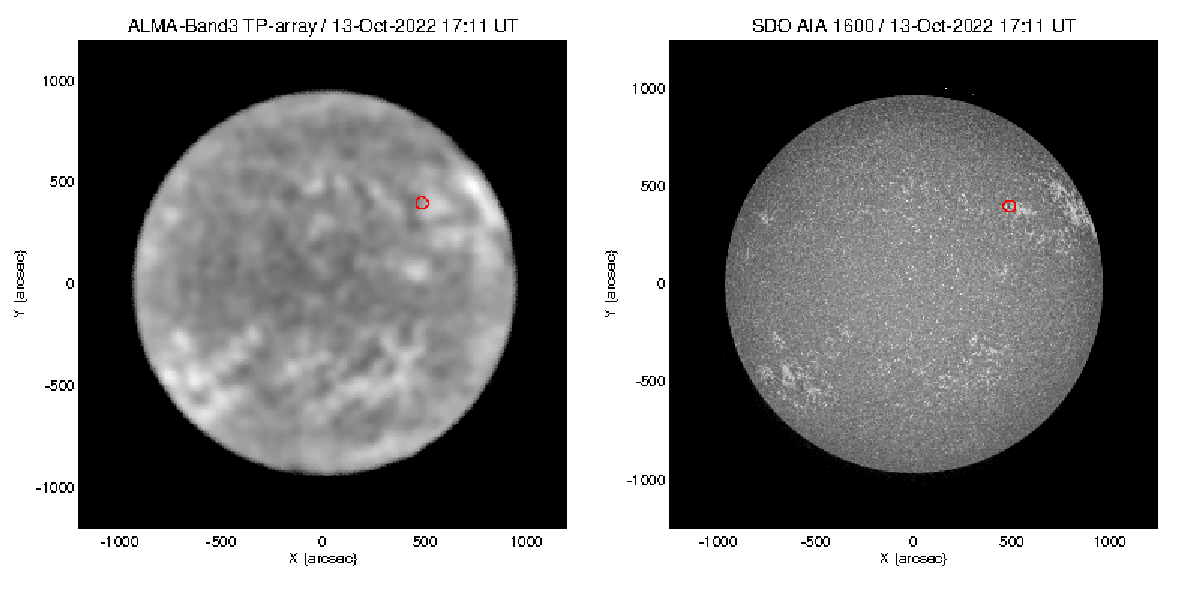}}
\caption{The full Sun maps obtained with the ALMA-Band3 (Left) and AIA 1600\AA~band (Right) at 17:11 UT, 13 October 2022. The red circles in both maps indicate the field of view of the images of Field ID \#0.} \label{fig:FullSun}
\end{figure}

Based on the suggestive results of the observation in May 2022, we designed a more comprehensive test observation with multi-pointings to 1) detect solar Stokes-V signals from fully calibrated data; 2) evaluate the effects of off-axis instrumental polarization. The observation with Band 3 ($f_{LO1}$: 100 GHz) was executed on 13 October 2022. The antenna configuration was C-3, and the correlator was used in the Time Division Mode (TDM). The total observing duration of the observation is 4.3 hours, including the three Execution Blocks (EBs). However, we were unable to use the third EB because the elevation of the Sun was too low, causing antennas to significantly shadow one another.  The datasets are released as Scientific Verification (SV) data (ADS/JAO.ALMA\#2011.0.00011.E), but it includes only 1st and 2nd EBs (1st EB: uid://A002/Xff99e1/X5b5, 2nd EB: uid://A002/Xff99e1/X66bc). The observing period of each EB was 15:11:48 -- 17:19:47~UT for 1st EB and 17:21:14 -- 18:47:37~UT for 2nd EB. While these interferometric observations were carried out with the heterogeneous array constructed with 7~m and 12~m antennas, we only analyzed the data obtained from the 12m antennas for reasons discussed in \S\ref{subsec:APCS}. The SV data also includes data only from the 12-m array. During the interferometric observations, we also obtained several full disk maps of the Sun using fast-scanning techniques \citep{2017SoPh..292...88W}. In this paper, we use the full-disk map of the Sun obtained with the TP-array from 17:08:32 -- 17:23:15~UT (uid://A002/Xff99e1/X666b). The TP data are also included in the SV data. 

\begin{figure} 
\centerline{\includegraphics[width=0.5\textwidth,clip=]{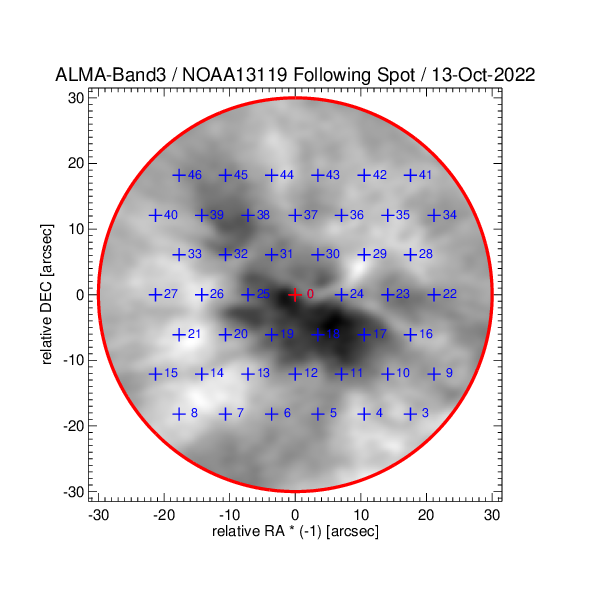}}
\caption{The synthesized Stokes~I image of FieldID \#0. The blue and red plus marks and number indicate the offset and Field ID of each pointing. The red circle indicates the FWHM of the primary beam. }\label{fig:Pointings}
\end{figure}

The target of October's observation was the following sunspot in the active region NOAA 13119 indicated by the red circle in Figure \ref{fig:FullSun}. We carried out a dense 45-pointing observation using the MOSAIC function with pointings spaced $7"$ apart. The offset of each pointing relative to the map center (red plus symbol) is indicated blue pluses in Figure \ref{fig:Pointings}.

During the observation, we observed 3C279 as the bandpass and polarization calibrators and J1256-0547 as the phase calibrator. The observing sequence for solar polarization calibration is the same as that used for non-solar celestial objects, except for the treatment for measuring the antenna temperature of the Sun described in \cite{2017SoPh..292...87S}. The MD-mode was used for the entire observation. We carried out the calibration of the dataset in two steps. The first step is to calibrate the parallel-hand data (XX and YY data) using the same calibration procedures that have been employed since 2016. The second step is to calibrate both the cross-hand (XY and YX) and parallel-hand data; i.e., infer the X-Y offset and the D-terms. Polarization calibration is the same as described in the ALMA Technical Handbook \citep{THBCy9}.   

\begin{figure}
\centerline{\includegraphics[width=1\textwidth,clip=]{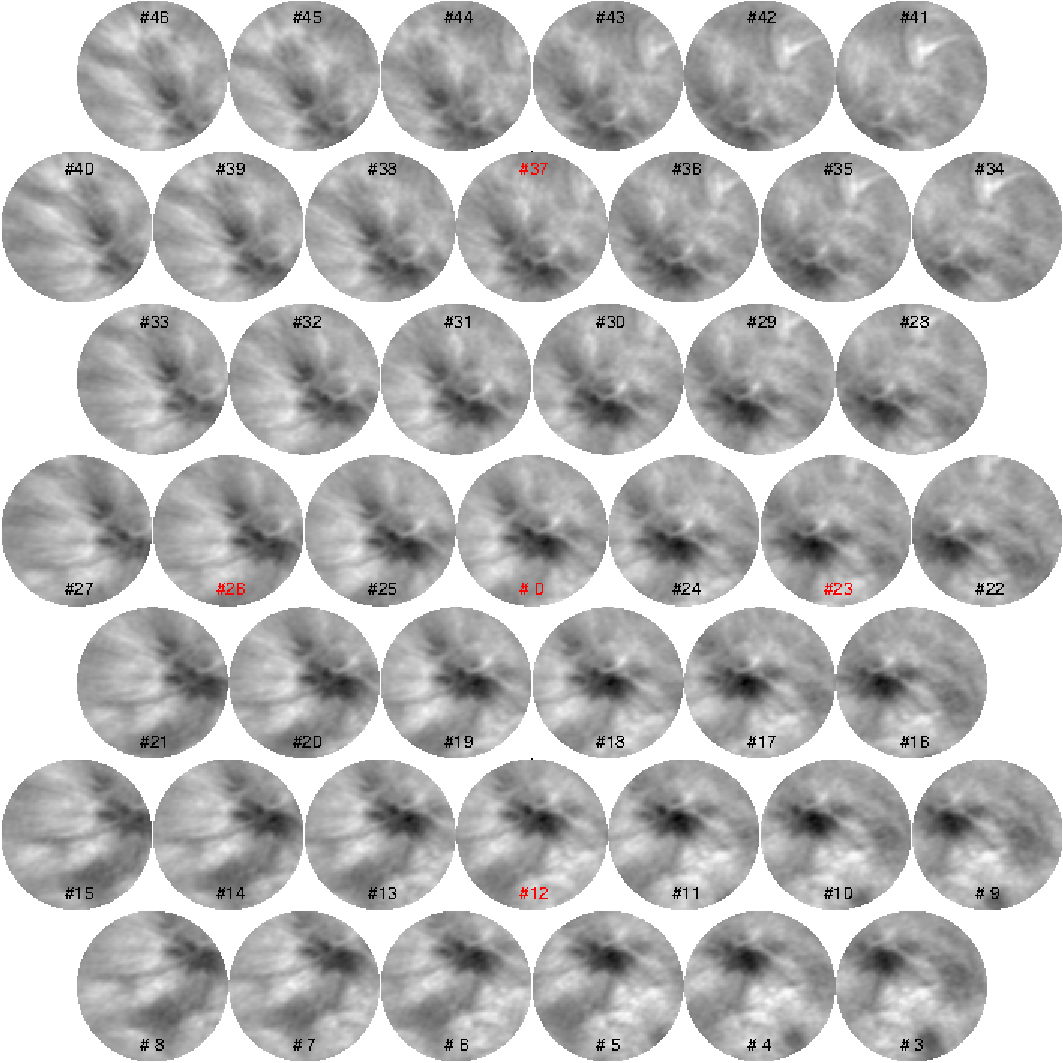}}
\caption{The Stokes-I maps of all fields synthesized from the 12m-array data. The number indicates the Field ID. The red Field IDs indicate the fields shown in Figure \ref{fig:field0}. The images are in the geocentric RA-DEC frame}\label{fig:Stokes-I}
\end{figure}

As a check of the polarization calibration using all antennas in MD mode, we calculated the degree of linear polarization and the electric vector position angle of the polarization calibrator, 3C279. The results are $9.534\pm0.008$\% and $16.85\pm0.03$ degrees, respectively. The ALMA observatory routinely monitors calibrator sources and records their properties in a database (AMAPOLA). Using this resource, the linear polarization degree and polarization angle of 3C279 were $10.0\pm1.3$\% and $12.2\pm7.3$ degrees, respectively on 22 September, and $9.9\pm2.7$\%, $21.1\pm3.8$ degrees on 22 October. Given the uncertainties, our results are consistent with the database, and the polarization calibration works well.

\begin{figure} 
\centerline{\includegraphics[width=1\textwidth,clip=]{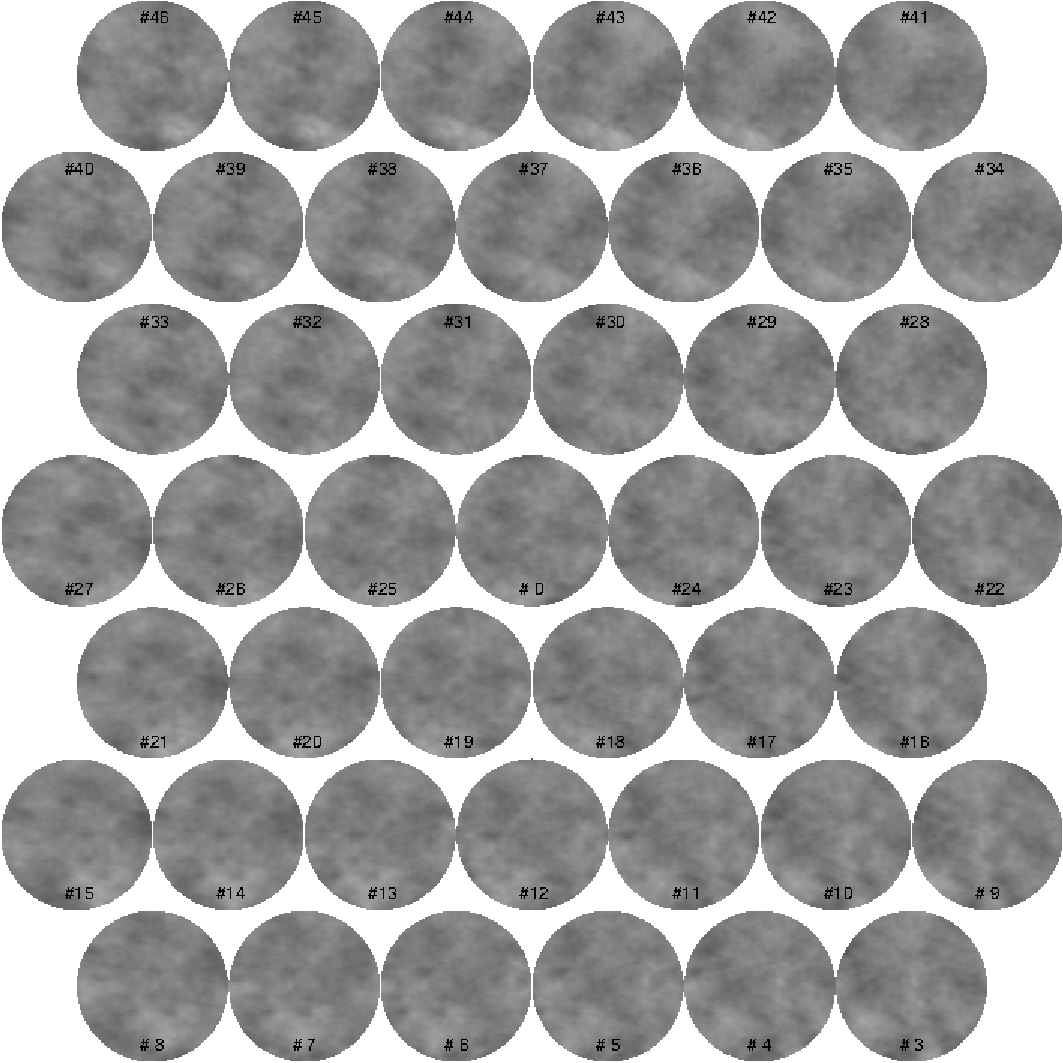}}
\caption{The Stokes~Q maps of all fields synthesized from the 12m-array data. The gray scale indicates from -2 Jy/beam to 2 Jy/beam.}\label{fig:Stokes-Q}
\end{figure}

\begin{figure} 
\centerline{\includegraphics[width=1\textwidth,clip=]{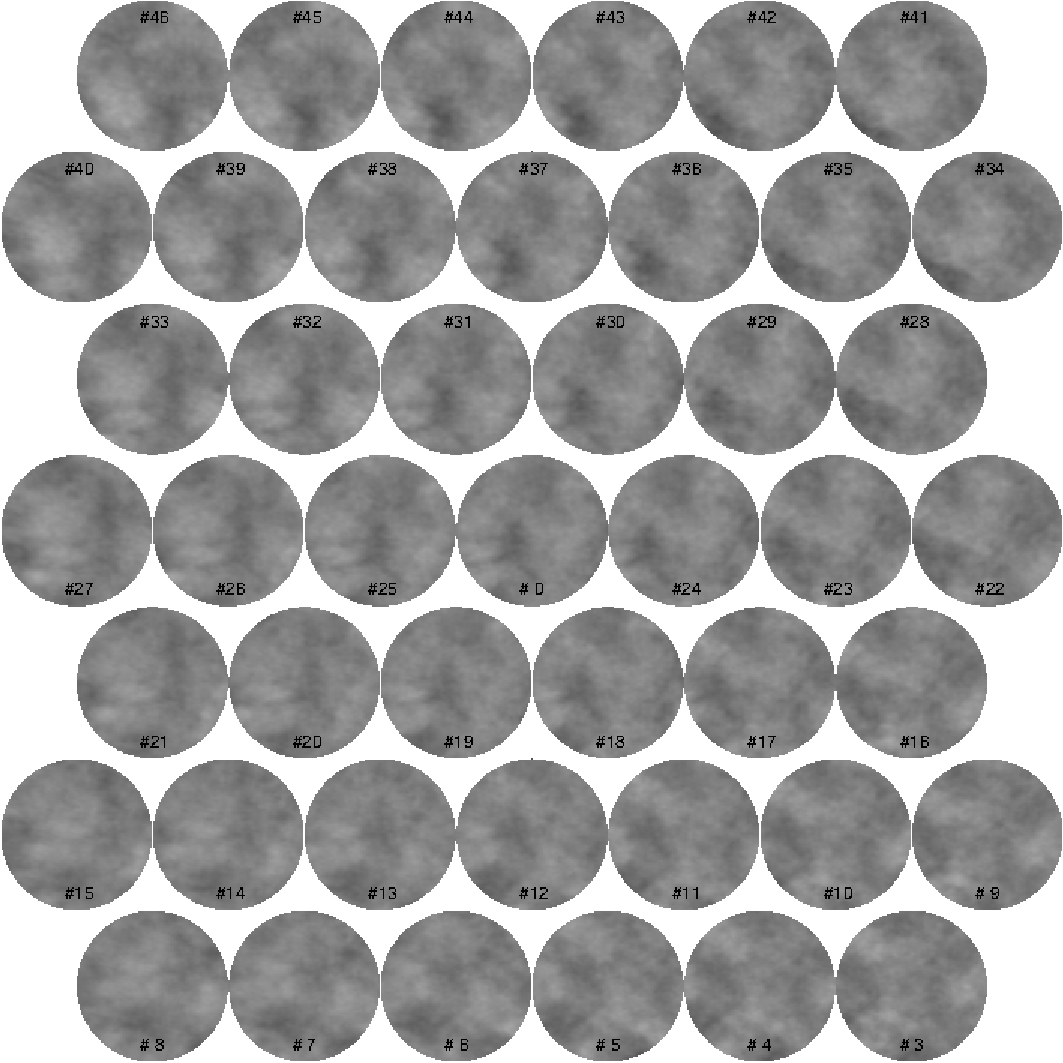}}
\caption{The Stokes~U maps of all fields synthesized from the 12m-array data. The gray scale indicates from -2 Jy/beam to 2 Jy/beam.}\label{fig:Stokes-U}
\end{figure}

\begin{figure} 
\centerline{\includegraphics[width=1\textwidth,clip=]{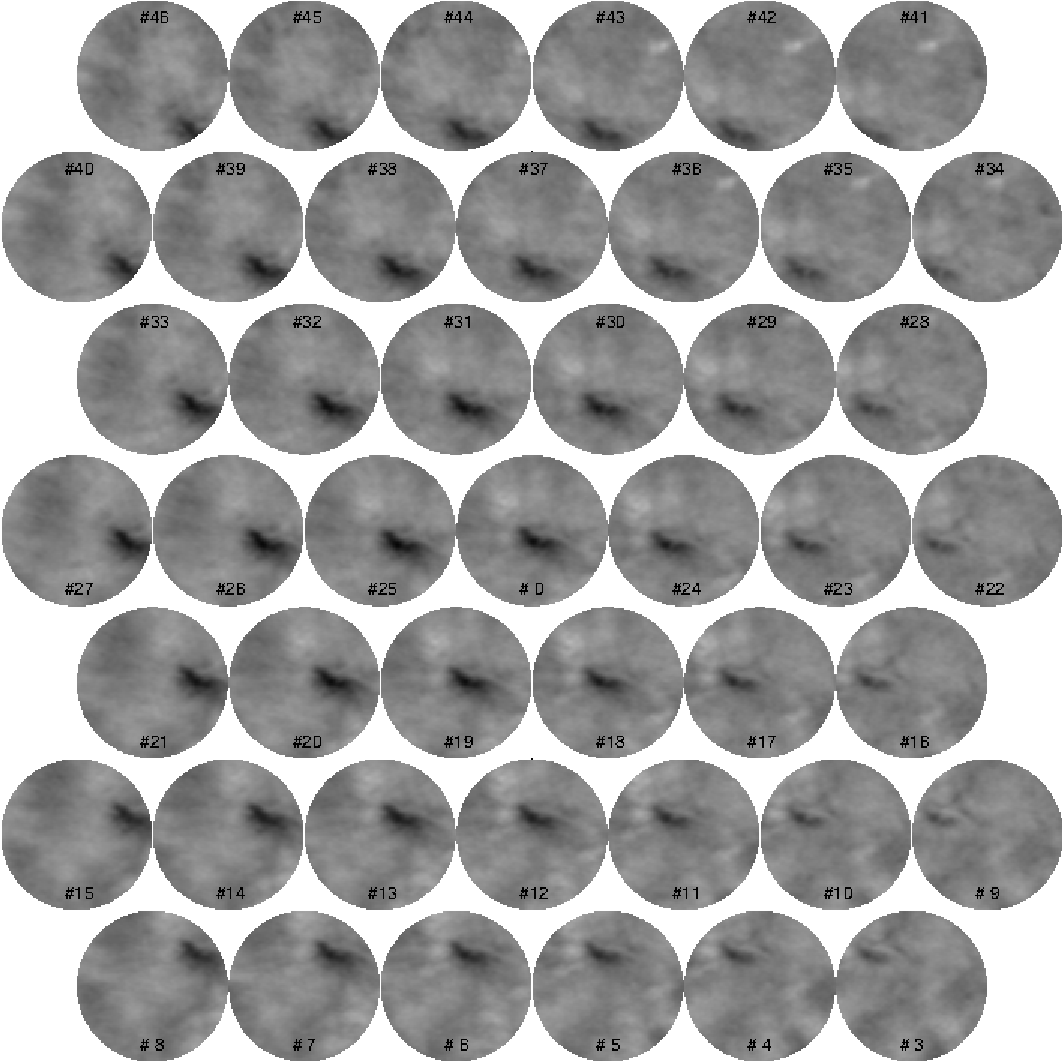}}
\caption{The Stokes~V maps of all fields synthesized from the 12m-array data. The  gray scale indicates from -2 Jy/beam to 2 Jy/beam.}\label{fig:Stokes-V}
\end{figure}

\subsubsection{Synthesized Full-Stokes Maps of the Sunspot}

We synthesized 3~mm maps in all four Stokes parameters I, Q, U, and V, for each pointing from the calibrated solar data. For the image synthesis, we use all scans of the sunspot for which the total integration time for each pointing is 67 seconds on average. The synthesized beam size is 2.06 arc-seconds for the major axis and 1.44 arc-seconds for the minor axis. No self-calibration was performed. 

In the Stokes~I maps (Figure \ref{fig:Stokes-I}), we can see the structures related to the sunspot, and we clearly identify the Stokes~V signals from above the sunspot in the Stokes~V maps (Figure \ref{fig:Stokes-V}).  Correction of the primary beam response has not been applied to these maps and so the  signals become weaker when the source approaches the edge of the field of view. Nevertheless, the Stokes~V signal is obvious in all pointings, even when it is located near the edge of the field of view. In contrast, we see no obvious signals in the Stokes~Q and U maps (Figure \ref{fig:Stokes-Q}, \ref{fig:Stokes-U}), as expected.

\begin{figure} 
\centerline{\includegraphics[width=1\textwidth,clip=]{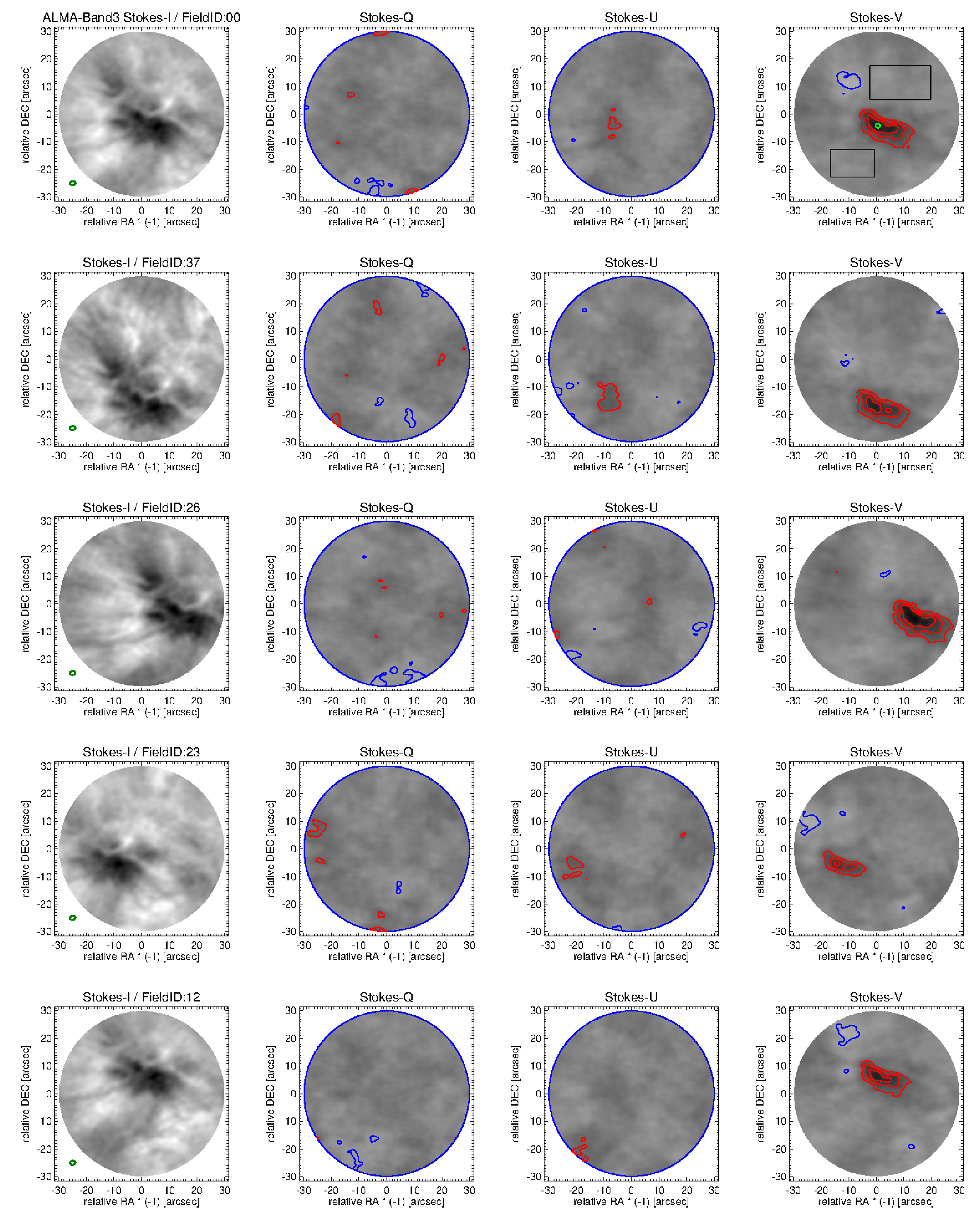}}
\caption{The Stokes~I, Q, U, and V maps (from Left) of Field ID \#0, 37, 26, 23, and 12 (from top). The green ellipse at the lower-left corner in the Stokes-I map shows the FWHM of the synthesized beam. The blue and red contours show the 3, 5, and 7 $\times$ RMS of each map (blue: positive, and red: negative). The light-green contour indicates 10 $\times$ RMS level of the Stokes~V map of Field ID \#0.}\label{fig:field0}
\end{figure}

\begin{figure} 
\centerline{\includegraphics[width=1\textwidth,clip=]{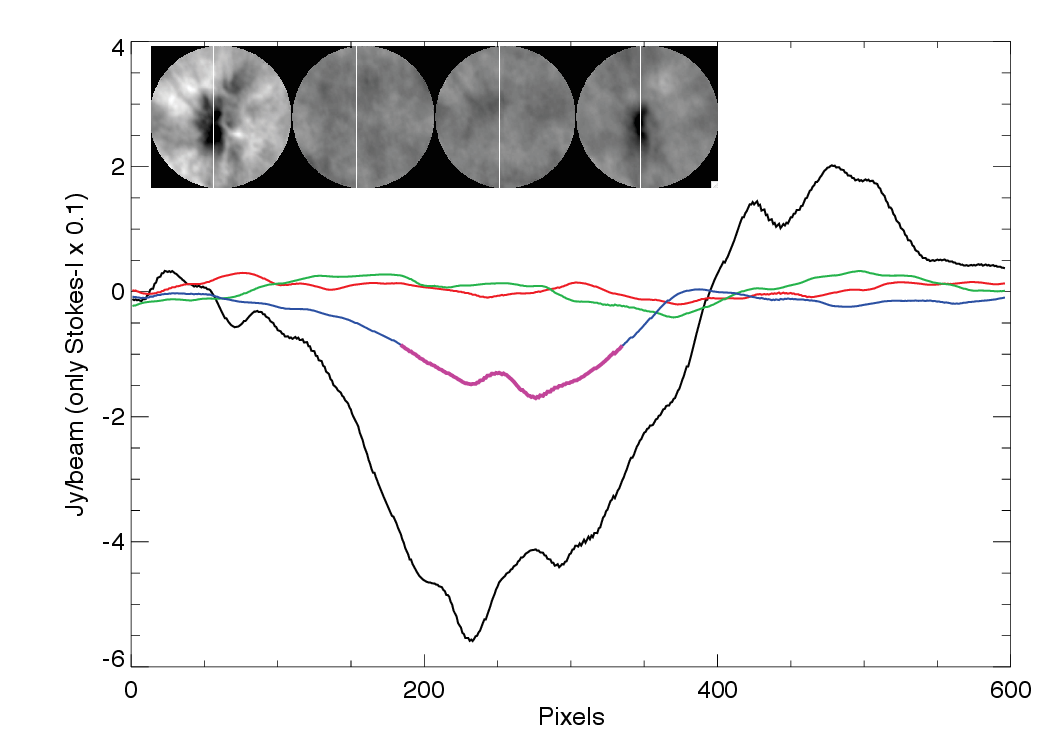}}
\caption{The spatial variation of Stokes parameters. Upper left panels: the Stokes~I, Q, U, and V maps of FieldID\#0 rotated 67 degree clockwise from the RA-Dec frame. Plots: The Stokes parameters along white lines in the upper left panels. Black: Stokes~I $\times$ 0.1, Red: Stokes~Q, Green: Stokes~U, Blue \& Purple: Stokes~V. The purple line shows the values are greater than 5 $\sigma$ level.}\label{fig:StkProf}
\end{figure}

Figure \ref{fig:field0} shows the Stokes~I, Q, U, and V maps of FieldID \#0, 37, 26, 23, and 12, the center, upper, left, right, and lower fields of the dense mosaic, and Figure \ref{fig:StkProf} is the spatial profile of the Stokes parameters through the center of the sunspot in FieldID \#0.  From the plot, the minimum value in the Stokes~I map is about -55 Jy/beam.  We note that the maps presented here are the result of the interferometric observation by the 12-m array. The background Sun has been resolved out by the array and the maps represent departures from the mean. For this reason, cooler regions -- the sunspot -- appear as negative flux density. We discuss the issue further in \S\ref{subsec:cpd}.

Usually, the significance of features in synthesized images is evaluated in comparison to the noise level using the root-mean-square (RMS) of the area where there is no significant source. We use the method for Stokes~Q, U, and V sources, and the RMSs of Stokes~Q and U maps are calculated from the flux density in the whole field of view. For Stokes~V maps, we use the areas indicated by the black boxes in the Stokes-V image of FieldID \#0 in Figure \ref{fig:field0} . The RMS of the Stokes~Q, U, and V maps is 0.14 Jy/beam, 0.14 Jy/beam, and 0.17 Jy/beam, respectively, which we adopt as the $1\sigma$ levels of the maps. 

The red (negative)  and blue (positive) contours in the Stokes~Q, U, and V maps (Figure \ref{fig:field0}) indicated 3, 5, and 7 $\sigma$ levels. Although all values in the Stokes~Q and U maps should be near 0, there are structures with values larger than 3$\sigma$ levels. Some of them are located near the field center. The structures might be caused by calibration errors or off-axis effects (see next section). In the absence of corrections for off axis effects we recommend using polarized sources with over 5-sigma levels for science.

\subsection{Assessment of Off-axis Polarization}\label{subsec:AssOff}

Telescopes with asymmetric optics like the ALMA and the Jansky VLA suffer from off-axis errors that affect observations of linearly and circularly polarized source emission \citep{1973ITAP...21..339C}.  For the Stokes parameters relevant to linearly polarized radiation (Q and U), the off-axis effects are referred to as ``beam squash" whereas for Stokes V the effect is referred to as ``beam squint" \citep{2001PASP..113.1247H}.  The impact of these off-axis effects is to induce an apparent polarization in the emission.  Recent characterizations of ALMA off-axis polarization are those of \citet{2020PASP..132i4501H} and \citet{ALMA-SPP}. These effects are independent of whether the antenna feeds are native linear (ALMA) or native circular (Jansky VLA). 

Figure \ref{fig:field0} shows strong emission in Stokes I and Stokes V in all pointings. The Stokes V emission is highly correlated with the Stokes I emission. On the other hand, the Q and U maps show no strong sources. However, there are areas in each of the Q and U maps that exceed 3-sigma. These are uncorrelated with emission in Stokes I and V. We attribute them to residual calibration errors and/or beam squash \citep[see][]{2020PASP..132i4501H}. There is no sign of the effects of beam squash in the pointing centered on the source (Figure \ref{fig:StkProf}). We therefore direct our attention to beam squint, which is relevant to Stokes V.

The antenna response function, referred to as the primary beam, is nearly Gaussian. It multiplies the distribution of millimeter-wavelength brightness on the sky. The field of view of the antenna is often taken to be the full width at half maximum of the Gaussian response function. For pedagogical reasons, it is convenient to consider squint in the context of orthogonal circularly polarized feeds; i.e., one feed receiving right-handed circular polarized (RCP) radiation and the other receiving left-handed circularly polarized (LCP) radiation. Beam squint manifests as a small effective offset of each beam in equal and opposite directions from the optical axis of the antenna. Following \citet{vla180}, suppose that the RCP and LCP primary beam responses are identical Gaussians with a half power beam width of $\theta_{FWHM}$, each offset in equal and opposite directions by $\theta_s=\sqrt{x_\circ^2+y_\circ^2}$. We can write the normalized antenna response to the two polarization as follows:

\begin{equation}
{\rm RCP ~beam}: P_{R}(x,y) = R_\circ e^{-\alpha^2[(x-x_{0})^{2}+(y-y_{0})^{2}]}  
\end{equation}
\begin{equation}
{\rm LCP ~beam}: P_{L}(x,y) = L_\circ e^{-\alpha^2[(x+x_{0})^{2}+(y+y_{0})^{2}]}
\end{equation}
where $\alpha^2 = 4\log 2/\theta_{FWHM}^2$. For small squint offsets $(x_\circ,y_\circ)$, $R_\circ \approx L_\circ \equiv A_\circ$, Eqns.~(1) and (2) may be written
\begin{equation}
P_{R}(x,y) \approx A_\circ e^{-\alpha^2(x^{2}+y^{2})} [1+2\alpha^{2}(xx_\circ+yy_\circ)] = A_\circ e^{-\alpha^2(x^{2}+y^{2})} [1+g(x,y)]
\end{equation}
\begin{equation}
P_{L}(x,y) \approx A_\circ e^{-\alpha^2(x^{2}+y^{2})} [1-2\alpha^{2}(xx_\circ+yy_\circ)] = A_\circ e^{-\alpha^2(x^{2}+y^{2})} [1-g(x,y)]
\end{equation}

\noindent where $g(x,y)=2\alpha^2(xx_\circ+yy_\circ)$. The beam response for Stokes I and V are then $P_I=(P_R+P_L)/2$ and $P_V=(P_R-P_L)/2$ from which it follows that the antenna response to Stokes-I and Stokes-V is

\begin{equation}
P_{I}(x,y) \approx A_\circ e^{-\alpha^2(x^{2}+y^{2})} 
\end{equation}
\begin{equation}
P_{V}(x,y) \approx g(x,y)A_\circ e^{-\alpha^2(x^{2}+y^{2})} 
\end{equation}

\noindent Consider a point source with an on-axis Stokes-I flux density $I_{on}$ and a Stokes-V flux density of $V_{on}$, and off-axis values of $I_{off}$ and $V_{off}$ respectively.  Using Eqns.~(11) and (12) we find

\begin{eqnarray}
V_{off}(x,y)=V_{on}e^{-\alpha^2(x^2+y^2)}+g(x,y)I_{on}e^{-\alpha^2(x^2+y^2)}
\end{eqnarray}

\noindent Noting that $I_{off}/I_{on}=\exp[-\alpha^2(x^2+y^2)]$, Eqn.~15 yields, in analogy to expressions given by \citet{2020PASP..132i4501H} for the fractional degree of linear polarization induced by beam squash, the fractional degree of circular polarization induced by beam squint:

\begin{equation}
\delta \rho^c(x,y)=\frac{V_{off}(x,y)}{I_{off}(x,y)}-\frac{V_{on}}{I_{on}} = g(x,y)
\end{equation}

% We should also include delta-V to compare with Hull's Fig. 3

\noindent We note that, even for an unpolarized source such that $V_{on}=0$, beam squint induces polarization when emission is observed off-axis. 

\citet{2020PASP..132i4501H} used dense mosaic imaging of a point-like calibrator source with known polarization properties (3C279) to evaluate the off-axis polarization properties of ALMA antennas. In contrast, we performed a dense mosaic of a sunspot in an active region, as described in \S3.2.  The sunspot appears to have a significant Stokes-V signal. Given that the source was observed using 45 offset pointings, we can use the observations to characterize the beam squint response in analogy to Hull et al. However,  to do so, we must address two issues. First, the Stokes~V source above the sunspot is not a point source. We therefore selected a region where the Stokes~V signal is $>10\sigma$ in FieldID\#0, the central field of the mosaic (the green contour in the Stokes-V image of Field ID \#0 in Figure \ref{fig:field0}) and use the averaged flux density of the source as a point-like Stokes-V signal, allowing us to estimate $V_{off}(x,y)$ and $V_{on}$. Second, as an interferometer, ALMA resolves out the background Sun, measuring only variations in flux density relative to the background. Hence, $I_{off}(x,y)$ and $I_{on}$ have negative values in the present case because the sunspot is darker than the mean background. We therefore use the absolute values of $I_{off}(x,y)$ and $I_{on}$ to calculate $\delta V$. 

\begin{figure} 
\includegraphics[width=0.5\textwidth,clip=]{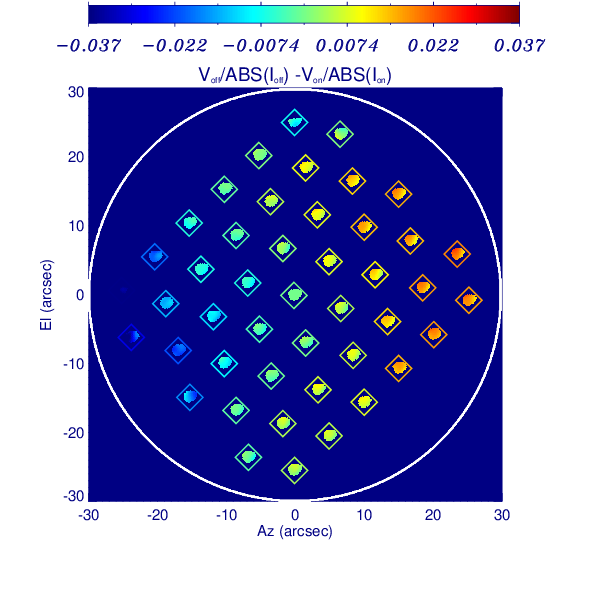}
\includegraphics[width=0.5\textwidth,clip=]{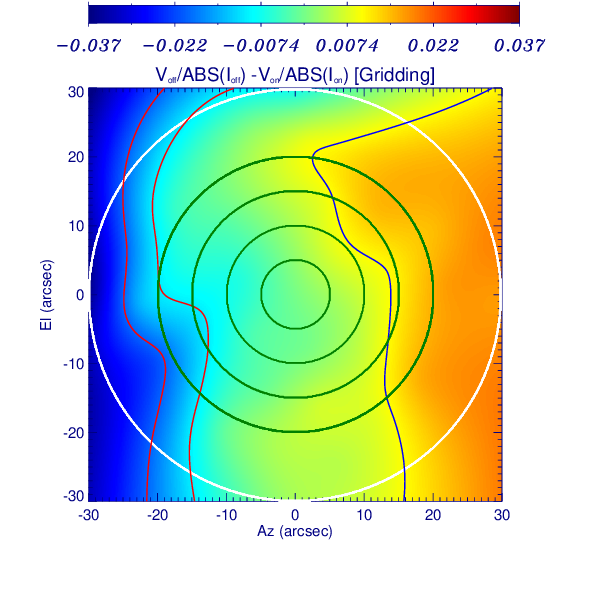}
\caption{The $\delta \rho^c$ map based on ALMA observations of a sunspot. In the right panel, the green contours indicate the distance from the center of the field of view, 5", 10", 15" and 20", the white circle indicates the field of view of an ALMA antenna in Band 3 (60"), and the red (negative) and blue (positive) contours indicate $\delta \rho^c$ of 1 and 2\%.}\label{fig:dVdata}
\end{figure}

To create the $\delta \rho^c$ map, then, we first made a mask based on the compact high signal-to-noise Stokes-V area in FieldID\#0 described above, and applied it to all offset fields and $\delta\rho^c$ was computed. The result is shown in the left panel of Figure \ref{fig:dVdata}. The color of each patch, corresponding to the masked area in each offset pointing, indicates the $\delta V$ distribution within the masked area. The color of the diamond surrounding the masked patches indicates the averaged values of $\delta\rho^c$. Since the distribution of $\delta\rho^c$ depends on the coordinate system of an antenna, the maps in Figure \ref{fig:dVdata} is displayed on the Az-El coordinate system. In the left panel of Figure \ref{fig:dVdata} the data in the right panel have been smoothed and interpolated to form a map of $\delta\rho^c$. We see that the result is consistent with expectations: that the off-axis polarization induced by beam squint shows an approximately linear gradient across the field of view of the 12m antennas. The value is small, no more than $\sim 2.5\%$ across the primary beam.

\subsection{Mitigating Beam Squint in Solar Polarization Maps}\label{subsec:migoff} 

We have shown that the fractional circular polarization induced by beam squint in a solar map conforms to expectations. We now consider the question of how to correct measurements of Stokes-V with ALMA for the effects of beam squint, or at least substantially mitigate them. Consider a source for which the intrinsic brightness distribution -- the ``true" brightness distribution -- in each sense of circular polarization is $RR(x,y)$ and $LL(x,y)$. The ``true" distributions of flux density in Stokes I and V are then $I_{true}(x,y)=[RR(x,y)+LL(x,y)]/2$ and $V_{true}(x,y)=[RR(x,y)-LL(x,y)]/2$, respectively. The observed distribution of flux in Stokes I is 

\begin{eqnarray}
I_{obs}(x,y)&=&\frac{RR(x,y)P_R(x,y)+LL(x,y)P_L(x,y)}{2}\\
&=& [I_{true}(x,y)+g(x,y)V_{true}(x,y)]e^{-\alpha^2(x^2+y^2)}  
\end{eqnarray}

\noindent Similarly, the observed distribution of flux density in Stokes V is

\begin{equation}
V_{obs}(x,y)= [V_{true}(x,y)+g(x,y)I_{true}(x,y)]e^{-\alpha^2(x^2+y^2)} 
\end{equation}

\noindent The degree of circular polarization is $\rho_{true}^c=V_{true}/I_{true}$ and the observed degree of circular polarization is therefore

\begin{equation}
\rho_{obs}^c(x,y)=\frac{V_{true}(x,y)+g(x,y)I_{true}(x,y)}{I_{true}(x,y)+g(x,y)V_{true}(x,y) }.
\end{equation}

\noindent Rearranging terms, the true degree of circular polarization is then

\begin{equation}
\rho_{true}^c(x,y)=\frac{V_{obs}(x,y)-g(x,y)I_{obs}(x,y)}{I_{obs}(x,y)-g(x,y)V_{obs}(x,y) }.
\end{equation}

\noindent Hence, if $g(x,y)$ is known and is the same for all antennas, it is possible to correct maps of the degree of polarization for beam squint explicitly -- in principle. In practice, there are a number of complications.

First, as discussed by \citet{ALMA-SPP}, the antennas in the ALMA 12-m array are not identical. They are of two types reflecting the fact that two manufacturers were used -- AEM and Vertex -- each producing 25 antennas for the 12-m array.  A program of holographic measurements of ALMA 12-m antennas performed by \citet{ALMA-SPP}. This study showed that TICRA\footnote{https://www.ticra.com/} models of the off-axis polarization response of the two antenna types were expected to be quite similar. However, Zernicke models fit to the average antenna illumination showed differences between the two antenna types. In addition, there were significant differences between antennas of the same type. The range of peak fractional leakage in Stokes V was 1-6\% for antennas of type DA and 2-8\% for antennas of type DV. Moreover, there was surprising variation in the position angle of the beam squint although the authors acknowledge that this may be due to an uncorrected residual phase difference between the two linearly polarized data channels -- the X-Y phase offset discussed in \S2.2. We note that \citet{2020PASP..132i4501H} take no explicit account of the variation in squint from antenna to antenna. Their maps of the off-axis polarization represent the average response of the antennas in aggregate in each Stokes parameter. Second, synthesis imaging observations such as those made by ALMA are typically performed over a finite range of time, during which the antenna polarization pattern rotates on the sky, smearing the effects of beam squint. This can be corrected in a piecewise fashion along the lines suggested by \citet{vla180}. A superior method that addresses both complications would be to use A-projection techniques to correct for off-axis polarization effects \citep{2013ApJ...770...91B}, but this approach requires accurate models of the polarization response of each antenna used in the array which are not currently available.

An alternative is to employ a pragmatic approach that corrects for beam squint approximately. The term $g(x,y)V_{obs}(x,y)$ in Eqn.~(21) is expected to be small across the field of view of any given antenna at wavelengths observed by ALMA. Even if each term is as large as 10\% of $I_{obs}$ the product is still only 1\% of $I_{obs}$. Neglecting the product, Eqn.~(21) simplifies to 

\begin{equation}
\rho_{true}^c(x,y)\approx\frac{V_{obs}(x,y)}{{I_{obs}(x,y)}}-g(x,y)=\rho_{obs}^c(x,y)-g(x,y)
\end{equation}

\noindent The function $g(x,y)$ represents the average response of all 12-m antennas used in the synthesis imaging. It can be deduced from the discrete fields of a mosaic observation of a circularly polarized solar source, as was done in \S3.3, or it can be deduced from observations of a calibrator source as was done by \citet{2020PASP..132i4501H}. 

Perhaps the simplest approach to mitigating the effects of beam squint is to performing mosaic imaging observations. Simulations have shown that beam squint effects are expected to be substantially reduced in larger format mosaics formed with Nyquist sampling. However, this mode has not yet been tested and validated with ALMA observations.

\subsection{The Importance of Total Power Data}\label{subsec:cpd} 

\begin{figure}
\centerline{\includegraphics[width=1\textwidth,clip=]{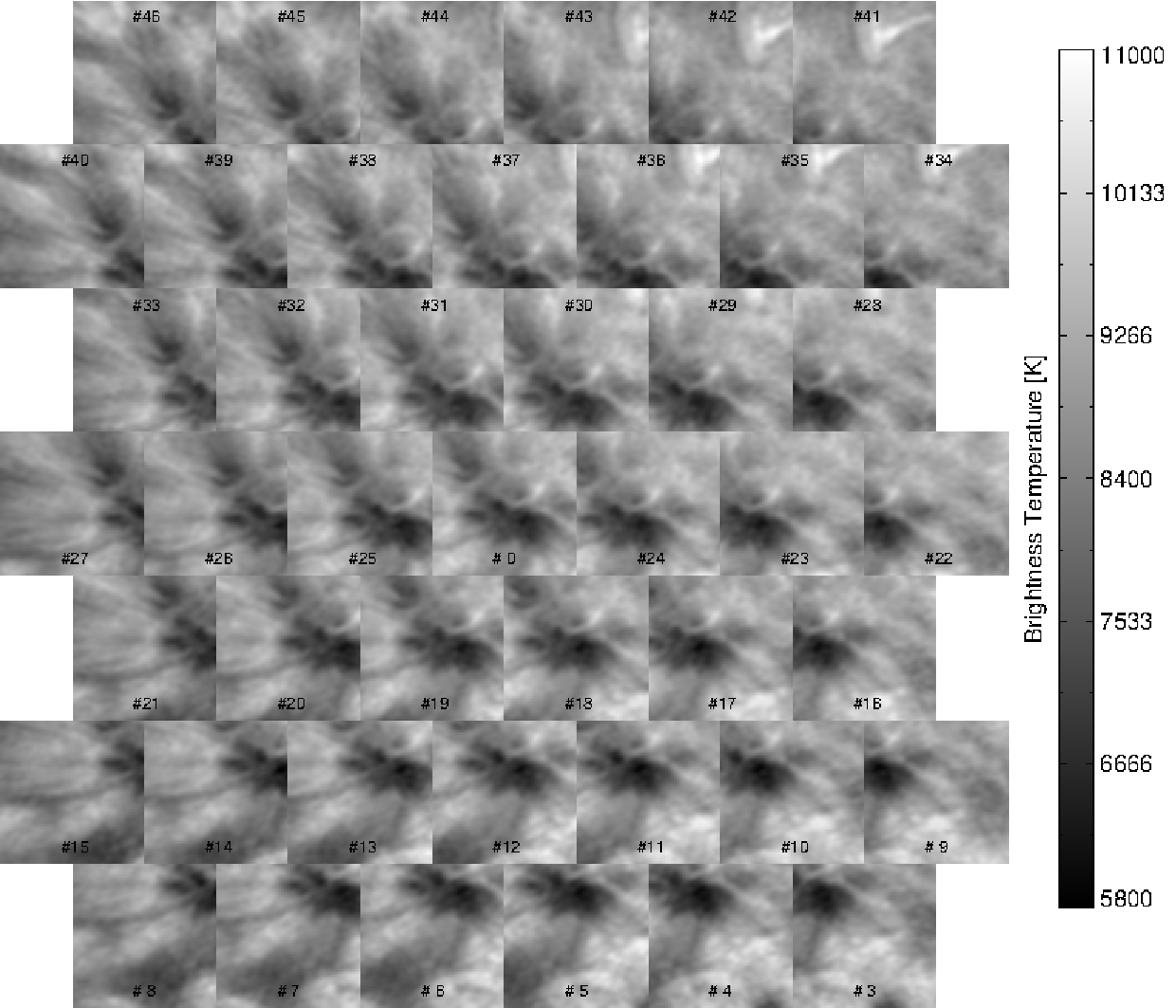}}
\caption{The Stokes-I maps of all fields after the feathering process. The number indicates the Field ID. Due to the feathering task can process only the square FoV, the image size and shape are changed from Figure \ref{fig:Stokes-I}. The images are on the RA-DEC frame}\label{fig:fth_K_multi}
\end{figure}

\begin{figure}
\centerline{\includegraphics[width=1\textwidth,clip=]{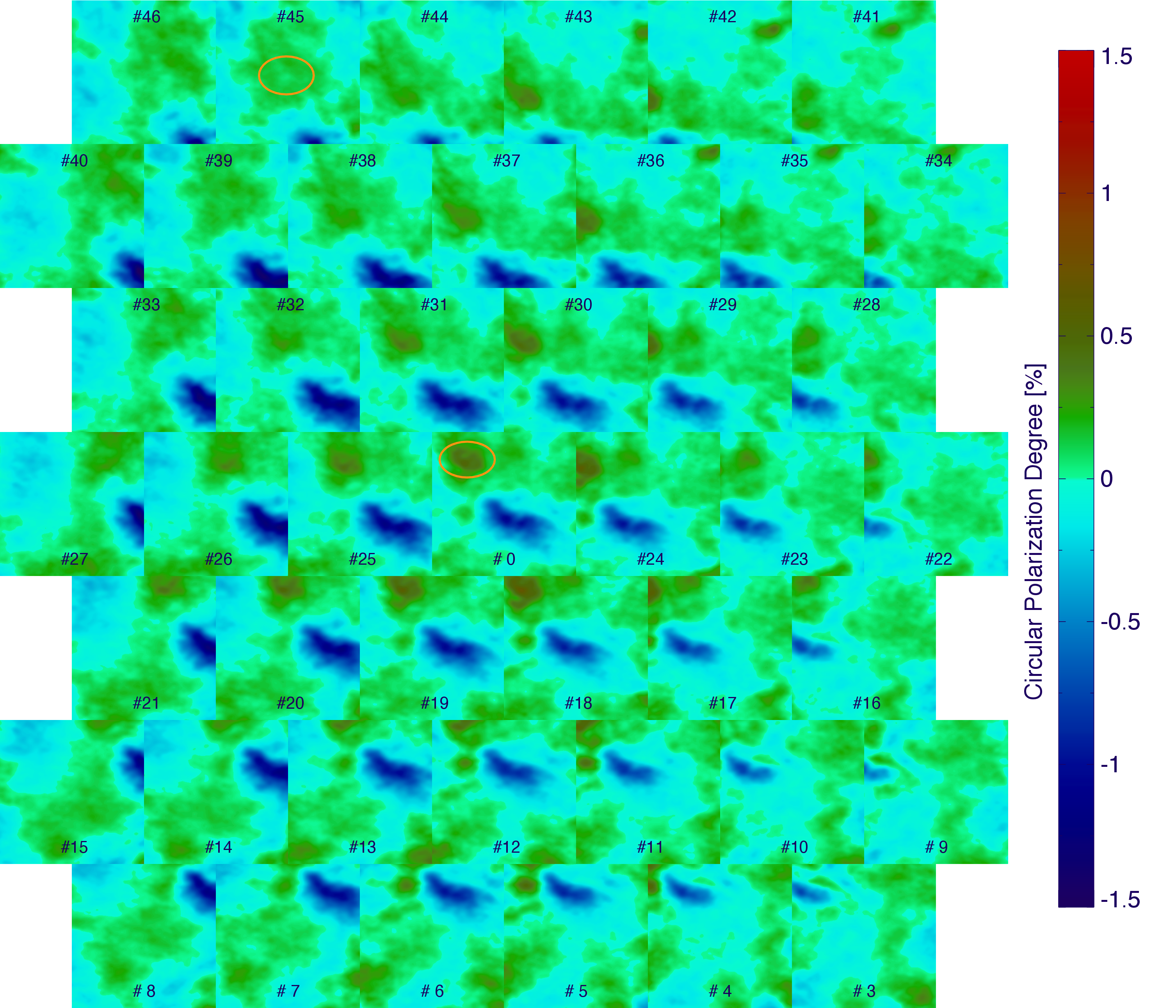}}
\caption{The circular polarization degree maps of all fields created from feathered Stokes~I map and synthesized Stokes~V map. The number indicates the Field ID. The images are on the RA-DEC frame}\label{fig:fth_cdp_multi}
\end{figure}

As discussed in \S2.1, interferometric observations with ALMA resolve out the background brightness distribution of the Sun. Accurate photometry in Stokes~I and accurate measurements of the degree of circular polarization $\rho_C=V/I$ require recovery of the background brightness distribution. This is done by performing low-resolution mapping of the Sun in total power \citep{2017SoPh..292...88W} and combining the low-resolution TP data with the high-resolution interferometric data using techniques such as feathering \citep{Obit41}. This has been done for each field of the dense mosaic presented here. Figure \ref{fig:fth_K_multi} shows each field in Stokes~I, calibrated in units of Kelvin. Figure \ref{fig:fth_cdp_multi} shows the percentage of the degree of circular polarization for the same fields.

\begin{figure} 
\centerline{\includegraphics[width=1\textwidth,clip=]{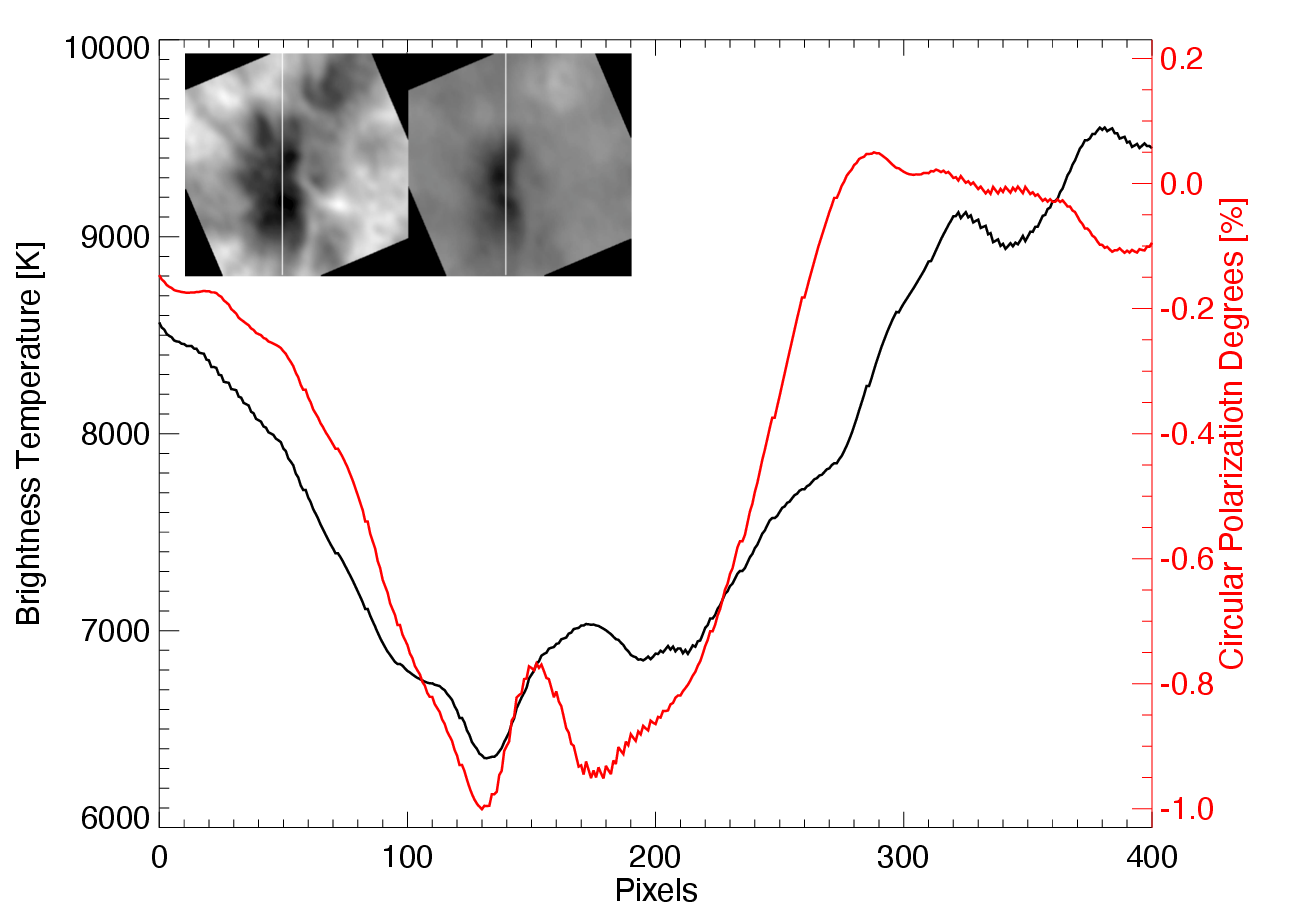}}
\caption{The spatial variation of brightness temperature and circular polarization degree. Upper left panels: the Stokes~I map with the feathering process, and circular polarization map rotated 67 degree clockwise from the RA-Dec frame.. Plots: The brightness temperature and circular polarization degree along white lines in the upper left panels. Black: Stokes~I, Red: Circular polarization degrees}\label{fig:CPDProf}
\end{figure}

Figure \ref{fig:CPDProf} shows the profile of Stokes~I brightness and circular polarization degree through the center of the sunspot. The plot shows that the circular polarization degree in the darkest region above the sunspot is -1 \%, which is similar to the prediction \citep{2017A&A...601A..43L, 2020FrASS...7...45L}. However, because TP data are only obtained in Stokes I, the degree of polarization should be regarded as a lower limit,  and the true value would become a little higher. 

% Can we estimate how much higher? Perhaps convolve the mosaic map made with the 45 pointings with the primary beam?

The other caution with solar polarization data is the effect of beam squint described in \S\ref{subsec:AssOff} and \ref{subsec:migoff} . We can see the positive Stokes~V source in the upper-left part of the circular polarization degree map of FieldID \#0 (orange circles in Figure \ref{fig:fth_cdp_multi}). The peak polarization degree of the source in FieldID \#0 is 0.5 \%. However, the polarization degree of the source is a few $\times$ 0.1 \% and is similar to that around it when the source is located at the center of the field of view (FieldID \#45). Thus, the positive Stokes~V source is an artificial source and is caused by beam squint. The source flux in the Stokes~V map of FieldID \#0 is 3 $\times$ RMS level (the Stokes-V image of Field ID \#0 in Figure \ref{fig:field0}). The artificial source in the circular polarization maps is another reason to recommend using sources with over 5 $\times$ RMS levels for science. 

\section{A New Observing Capability}\label{s:NewCap}

Based on the results of the test observations described above, the ALMA observatory began supporting solar polarization observations in ALMA Cycle10. In this section, we summarize the new observing mode compared to the previous solar observing modes and note the usage of the observing functionality based on the issues described in the previous section.

First, we list the properties of the solar polarization observations in ALMA Cycle10. The following properties are consistent with the ALMA Technical Handbook for Cycle 10, and we add some details of the functionality.

\begin{itemize}

\item Solar polarization observations may be performed with the 12m-array. The usable array configurations of the 12m-array are the same as for standard solar observations with Band3, i.e. C-1, C-2, C-3, and C-4 configurations. 7~m-antennas are not used. 

\item Only Band 3 receivers can be used for solar polarization observations at a wavelength of 3~mm. The spectral setting is the same as for the standard solar observations with Band 3. The frequency of the local oscillator \#1 is 100 GHz, and only the Time Division Mode (TDM), which is an ALMA's correlator mode for observing continuum emission, can be used.

\item The correlator outputs four cross-correlation data (XX, YY, XY, YX) with an integration time of 1~s. Four spectral windows are output, each with 64 channels. 

\item Scanning solar observations with the TP-array are performed simultaneously with the interferometric observations. However, the observations are exactly the same as standard solar observations; that is, only total intensity maps (Stokes I) are produced. PI can select full-Sun mapping or fast regional mapping for the TP-array observations.

\item By observing a polarization calibrator, the duty cycle of the target (solar) observations is reduced compared to standard solar observations. Figure \ref{fig:listobs} shows the sequence in an EB of a solar polarization observation, and the orange boxes in the lower part of the figure indicate the time ranges of observing the Sun in an EB. 

\begin{figure} 
\centerline{\includegraphics[width=0.95\textwidth,clip=]{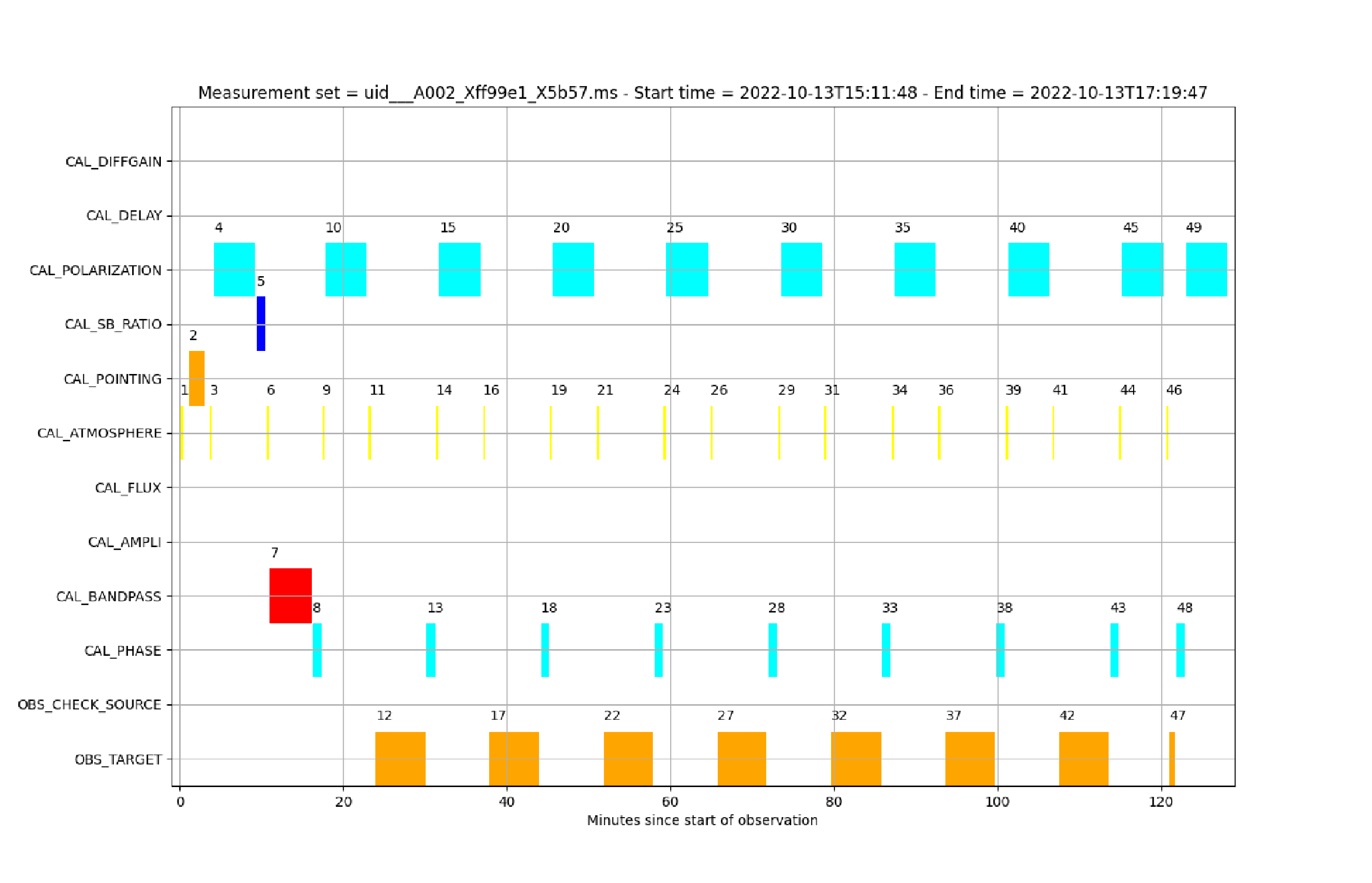}}
\caption{The structure of an Execution Block (EB) for a 3~mm solar polarization observation (uid://A002/Xff99e1/X5b57)}\label{fig:listobs}
\end{figure}

\item For polarization calibration, we need to know the dependence on the parallactic angle. Therefore, the total observation time for the current ALMA calibration method must be longer than 3 hours, even if the required on-source time is shorter than 3 hours. Thus, a Scheduling Block (maximum duration $\sim 90-120$ min) would be executed at least twice for one solar polarization observation.

\end{itemize}

\noindent The other components of a solar polarization observations are the same as for standard solar observations.

Predicting the coordinates of interesting solar targets a few days in advance to within a few arc seconds is not possible, in general, especially for transient phenomena like solar flares. Therefore, we can easily imagine that sometimes interesting structures appears near the edge of the field of view. In such cases, the beam squint would cause artificial Stokes V sources, as shown in Figure \ref{fig:fth_cdp_multi}, and should therefore be treated with caution. The effects of beam squint may be corrected or mitigated as described in \S3.4: i) explicit correction using Eqn.~(21), or; 2) approximate correction using Eqn.~(22). While both of these approaches require an estimate of $g(x,y)$, The data for estimating $g(x,y)$ described in \citet{2020PASP..132i4501H} is provided as the SV data (ADS/JAO.ALMA\#2011.0.00009.E) by the ALMA observatory.

\section{Conclusions}\label{s:Con}

\begin{figure} 
\centerline{\includegraphics[width=0.95\textwidth,clip=]{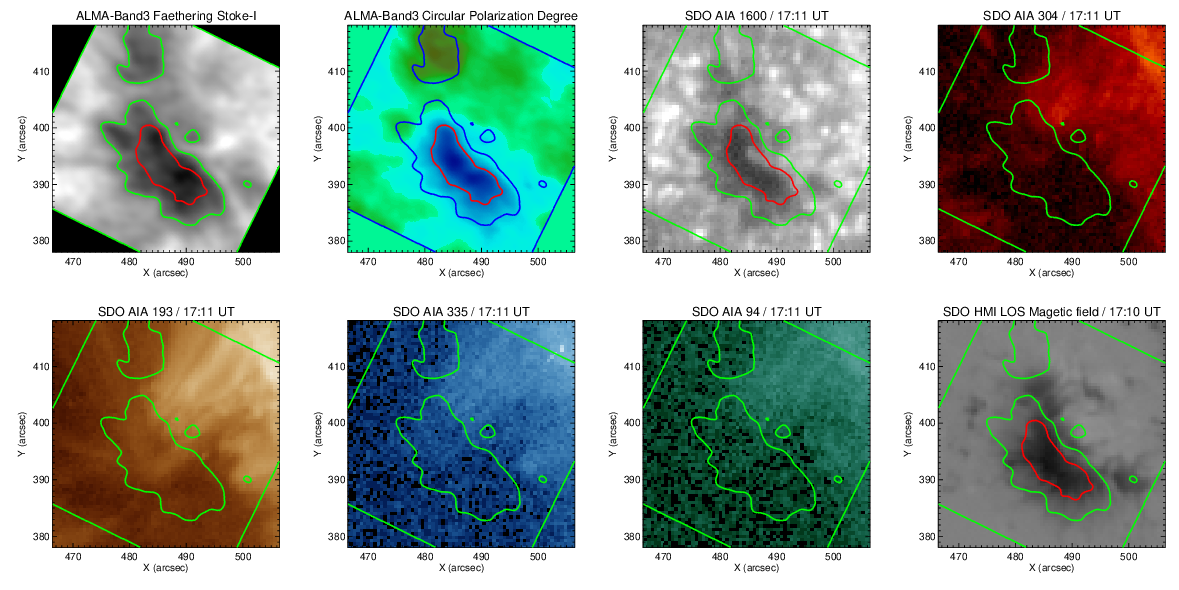}}
\caption{The maps obtained with the ALMA and SDO. From left; Upper panels: Stokes-I map with the feathering process (FieldID \#0), Circular polarization degree map of 100 GHz, AIA 1600 \AA band image. The primary beam correction is applied to the Stokes~I and circular polarization degree maps. Lower panel: the maps of AIA 193 \AA, 335 \AA, 94 \AA Bands and LOS magnetic field strength obtained with HMI/SDO. The green and blue contours show the Stokes~I map with the feathering process, and the red contours indicate the 5 $\times$ RMS level in the Stokes~V map (see Figure \ref{fig:field0}). The images are on the Heliocentric coordinate frame.}\label{fig:ALMA-SDO}
\end{figure}

Circular polarization measurements at millimeter wavelengths constrain the magnetic field environment in the upper solar atmosphere and is therefore an essential tool for advancing our knowledge of the Sun. Although there are some limitations and complexities in dealing with and understanding the datasets as described in this paper, solar polarization observations with ALMA will provide us with such critical information. As an example, we show the results of the co-alignment between ALMA and SDO images (Figure \ref{fig:ALMA-SDO}). Comparing the circular polarization degree map and the magnetogram obtained with HMI, the strong polarization signal comes from just above the strong photospheric magnetic fields of the sunspot. In the EUV images obtained with AIA, there is no significant structure with coronal temperatures above the polarization source. This is a predicted feature, but the relationship would be further investigated using the chromospheric magnetic fields obtained with ALMA. The circular polarization maps in this paper are synthesized from only 1 minute of integration. Thus, the time variance of the chromospheric magnetic fields would be an interesting topic based on ALMA observations. 

%%%%%%%%%%%%%%%%%%%%%%%%%%%%%%%%%%%%%%%%%%%%%%%%%%%%%%%%%%%%%%%%%%%%%%%%%%%
%% Acknowledgements
%

\begin{acks}
The ALMA solar-commissioning effort was supported by ALMA Development grants from NAOJ (for the East Asia contribution), NRAO (for the North American contribution), and ESO (for the European contribution). The help and cooperation of engineers, telescope operators, astronomers-on-duty, Extension and Optimization of Capabilities (EOC; Formerly Commissioning and Science Verification) team, and staff at the ALMA Operations Support Facility was crucial for the success of solar-commissioning campaigns in 2019 and 2022. We are grateful to the ALMA project for making solar observing with ALMA possible. This article makes use of the following ALMA data: ADS/JAO.ALMA\#2011.0.00011.E. ALMA is a partnership of ESO (representing its member states), NSF (USA), NINS (Japan), together with NRC (Canada), NSC, ASIAA (Taiwan), and KASI (Republic of Korea), in cooperation with the Republic of Chile. The Joint ALMA Observatory is operated by ESO, AUI/NRAO, and NAOJ. The National Radio Astronomy Observatory is a facility of the National Science Foundation operated under a cooperative agreement by Associated Universities, Inc. SDO is the first mission to be launched for NASA’s Living With a Star (LWS) Program. MS thanks the Joint Research Program of the Institute for Space-Earth Environmental Research (ISEE), Nagoya University, and Dr. Satoshi Masuda. 
\end{acks}

%%% %%%%%%%%%%%%%%%%%%%%%%%%%%%%%%%%%%%%%%%%%%%%%%%%%%%%%%%%%%%
%% Bibliography
%
% Using BibTeX
%
%\bibliographystyle{spr-mp-sola}
%\bibliography{main}  
%
% Without BibTeX 

%\end{article} 
\end{document}